\documentclass[a4paper,aps,prd,showpacs,showkeys,preprint] {revtex4}
\usepackage{amsfonts}
\usepackage{verbatim}
\usepackage{latexsym}
\usepackage{amsmath}
\usepackage{amssymb,bm}
\usepackage[dvips]{graphicx}
\usepackage{epsfig}        

\begin{document}

\title  {Topologically massive non-Abelian theory: superfield approach}

\author{S. Krishna$^{1}$}
\email{skrishna.bhu@gmail.com}
\author{A. Shukla$^{1}$}
\email{ashukla038@gmail.com}
\author{R. P. Malik$^{1,2}$}
\email{malik@bhu.ac.in}

\affiliation {$^{1}$Department of Physics, Centre of Advanced Studies,
 Banaras Hindu University, Varanasi - 221 005, (U. P.), India,\\
 $^{2}$DST Centre for Interdisciplinary Mathematical Sciences,
 Faculty of Science, Banaras Hindu University, Varanasi - 221 005, India}\

\begin{abstract}
We apply the well-established techniques of geometrical superfield approach to 
Becchi-Rouet-Stora-Tyutin (BRST) formalism in the 
context of four (3 + 1)-dimensional (4D) dynamical non-Abelian 2-form gauge theory by
exploiting its inherent ``scalar'' and ``vector'' gauge symmetry transformations and 
derive the corresponding off-shell nilpotent and 
absolutely anticommuting BRST and anti-BRST symmetry transformations. Our approach
leads to the derivation of {\it three} (anti-)BRST invariant Curci-Ferrari (CF)-type restrictions that are
found to be responsible for the absolute anticommutativity of the BRST and anti-BRST
symmetry transformations. We derive the coupled Lagrangian densities that respect the
(anti-)BRST symmetry transformations corresponding to the ``vector'' gauge transformations. 
We also capture the (anti-)BRST invariance of the
CF-type restrictions and  coupled Lagrangian densities within the framework
of our superfield approach. We obtain, furthermore, the off-shell nilpotent (anti-)BRST symmetry transformations
when the (anti-)BRST symmetry transformations corresponding to the ``scalar'' and ``vector'' gauge
symmetries are merged {\it together}. These off-shell nilpotent ``merged'' (anti-)BRST symmetry transformations 
are, however, found to be non-anticommuting in nature.
\end{abstract}

\pacs {11.15.Wx; 11.15.-q; 03.70.+k}

\keywords {Topologically massive 4D non-Abelian gauge theory; Curci-Ferrari type restrictions; nilpotency
and absolute anticommutativity; (anti-)BRST symmetry transformations;
superfield formulation }

\maketitle

\begin{center}
{\section{Introduction}}
\end{center}

During the past few years, there has been an upsurge of interest in the study of higher $p$-form
($p = 2,3,4,....$) gauge theories because of their relevance in the context of theory of (super) strings and
related extended objects (see, e.g. [1-3]). It has been found, furthermore, that the merging 
of 1-form gauge field and higher $p$-form ($p = 2, 3, 4...$) gauge fields has led to very interesting models of 
field theories which encompass in their folds rich mathematical
and physical structures. In particular, the coupling of 1-form and 2-form gauge fields
 has provided us models for the topologically massive gauge field 
theories in 4D. In recent years, there has been a renewed interest in the understanding of 4D massive 
topological gauge field theories of Abelian and non-Abelian variety [4-8] which provide an alternative to the 
celebrated Higgs mechanism that is responsible for generating masses for
the gauge particles and fermions of the standard model of particle physics.
It may be mentioned, in the context of our present endeavor, that the model under
 consideration addresses only the question of mass generation of the gauge field and
it does not shed light on any other off-shoots of the Higgs mechanism that play important roles 
in the standard model of high energy physics.

Despite many success stories, there are a couple of loop-holes in the physical foundations of the standard model
of high energy physics. For instance, the esoteric Higgs particles have not yet been observed experimentally and neutrinos
have been found to have mass by experimental techniques. These experimental observations have propelled physicists
to look for an alternative to the Higgs mechanism for the generation of masses for the gauge bosons and fermions
of the standard model of particle physics. In this context, it is pertinent to point
out that the 1-form gauge field acquires mass [4] when it is coupled with the 
antisymmetric tensor gauge field  $B_{\mu\nu}$ through the well-known 
topological term $B^{(2)} \wedge F^{(2)}$ where the 2-forms 
$B^{(2)} = \frac {1}{2!}\;(dx^\mu \wedge dx^\nu) B_{\mu\nu}$
and  $F^{(2)} = \frac {1}{2!}\;(dx^\mu \wedge dx^\nu)F_{\mu\nu}$ define the potential $B_{\mu\nu}$ 
and curvature tensor $ F_{\mu\nu} = \partial_\mu A_\nu - \partial_\nu A_\mu + i [A_\mu, A_\nu]$ corresponding 
to the 1-form $(A^{(1)} = dx^\mu A_\mu)$ gauge field $A_\mu$ of the 4D non-Abelian gauge theory, respectively. Thus, this 
field theory does provide a topological mass to the 1-form gauge boson without taking any recourse to the basic tenets of Higgs mechanism.
This observation is important in view of the fact that the Higgs particles have not yet been observed experimentally.

We have studied the 4D massive topological Abelian gauge theory 
within the framework of BRST formalism where the superfield technique
has played a very decisive role [9]. Its straightforward generalization to the 4D 
massive topological non-Abelian theory is non-trivial as it is plagued by 
some strong no-go theorems [10,12]. There are, however, a couple of models for the massive 
topological non-Abelian theory [5,13] which 
circumvent the severe strictures laid down by the above no-go theorems. 
In our present investigation, we focus on the dynamical non-Abelian
2-form gauge theory [13] and exploit its inherent ``scalar'' and ``vector'' gauge symmetry transformations within the framework of geometrical superfield formalism (see, e.g. [14-17]). In particular, we derive
the off-shell nilpotent and absolutely anticommuting
(anti-)BRST symmetry transformations corresponding to the ``scalar'' and ``vector'' gauge 
symmetries of the theory by exploiting the geometrical superfield formalism [14,15]. In our earlier work [18], we
have deduced the coupled Lagrangian densities that respect the (anti-)BRST symmetry 
transformations corresponding to the ``scalar'' gauge 
symmetry of the theory. In our present investigation, we derive the coupled Lagrangian densities that are 
found to be equivalent
and respect the (anti-)BRST symmetry transformations corresponding to the ``vector'' gauge 
transformations of the theory.

Our superfield formalism leads to the derivation of {\it three} (anti-)BRST invariant 
CF-type restrictions that enable us to
derive the equivalent coupled Lagrangian densities and lead us to
achieve the absolute anticommutativity property of the (anti-)BRST symmetries.
The emergence of the CF-type restrictions is an essential ingredient of the
application of our superfield formulation to any arbitrary $p$-form gauge theory.
Its deep connections with the concept of gerbes have been established in
our earlier works for the Abelian 2-form and 3-form gauge theories [19,20].
One of the key features of our superfield approach is the observation that the
ensuing CF-type restrictions are {\it always} (anti-)BRST invariant. In fact,
in our present endeavor, we have been able to capture the (anti-)BRST invariance
of the CF-type restrictions in the language of superfield formalism
itself  (see, Sec. VI below).

In our present paper, we {\it also} obtain the off-shell nilpotent (anti-)BRST symmetry transformations where the
(anti-)BRST symmetry transformations, corresponding to the ``scalar'' and ``vector'' gauge symmetries, are combined together. We christen these symmetries as the ``merged'' (anti-)BRST symmetries.
As it turns out, these ``merged'' (anti-)BRST symmetries, even though off-shell nilpotent, are found to be 
{\it not} absolutely anticommuting. As a result, they are {\it not} linearly independent and, thus, are not proper
in the true sense of the word. We derive, furthermore, the BRST and anti-BRST
invariant Lagrangian densities that respect the ``merged''
BRST and anti-BRST symmetries {\it separately} but they do not 
respect these symmetries {\it together} (because of the fact that the ``merged'' (anti-)BRST symmetries
are {\it not} absolutely anticommuting in nature). To obtain the absolute anticommutativity of the 
above (anti-)BRST symmetry transformations remains an open problem for us which we hope to address in the future
by exploiting the geometrical  superfield formalism.

Our present study is essential on the following grounds. First and foremost,
it is very important for us to generalize our earlier work on the topologically
massive Abelian gauge theory [9] to the case of non-Abelian theory. Second, 
the anti-BRST symmetry transformations have {\it not} been obtained in [7,13].
As a consequence, the requirement of anticommutativity property with the 
BRST transformations has remained an open problem. Third,
having derived the proper (anti-)BRST 
symmetry transformations corresponding to ``scalar'' gauge 
transformations [18], it is but natural for us to obtain the off-shell
nilpotent and absolutely anticommuting (anti-)BRST symmetry transformations
corresponding to the ``vector'' gauge  transformations by exploiting
the geometrical superfield formulation [14,15].
Fourth, it is interesting to observe that the {\it three} CF-type restrictions
emerge from our superfield approach that enable us to obtain absolutely anticommuting
(anti-)BRST symmetry transformations for the ``scalar'' and ``vector''
gauge symmetries of the theory. Finally, we obtain the ``merged'' (anti-)BRST
symmetry transformations from the above (anti-)BRST symmetries that are off-shell 
nilpotent but not absolutely anticommutating in nature.

The contents of our present paper are organized as follows. In Sec. II, to set
up the basic notations, we give a brief synopsis of the local ``scalar'' and 
``vector'' type gauge symmetry transformations that are present in the theory.
For our paper to be self-contained and to fix-up the supersymmetric notations, our
Sec. III deals with the derivation of off-shell nilpotent (anti-)BRST symmetry transformations
corresponding to the ``scalar'' gauge symmetry within the framework of superfield formulation [18].
In Sec. IV, we apply the superfield formalism to deduce the (anti-)BRST symmetry transformations corresponding to
the ``vector'' gauge symmetry transformations where the horizontality-type
restriction plays an important role. Our Sec. V is thoroughly devoted to the discussion of
(anti-)BRST invariance of the coupled Lagrangian
densities as well as the CF-type restrictions. These
invariances are captured within the framework of superfield formalism
in Sec. VI. Our Sec. VII deals with the off-shell nilpotent (anti-)BRST symmetries that correspond to
the ``scalar'' and ``vector'' gauge symmetries when they are merged {\it together}.
Finally, we make some concluding remarks and point out
a few future directions for investigations in Sec. VIII.\\

In our Appendix, we capture the no-go theorem, proposed in [10], within the framework of our present
geometrical superfield formalism.

{\bf{\it Conventions and notations:}} We follow here the conventions and notations 
such that the background 4D Minkowski spacetime manifold has the flat metric with signatures
(+1, -1, -1, -1) and the generators $T^a$ of the $SU(N)$ group obey the Lie algebra
$[T^a, T^b ] = i f^{abc} T^c$ with structure constants $f^{abc}$ (chosen to be
totally antisymmetric in indices $a, b$ and $c$ where $a, b, c....= 1, 2, ...N^2 -1$ ). 
In the algebraic space, we  adopt the notations: $(V \cdot W) = V^a W^a$ and $(V \times W)^a = f^{abc} V^b W^c$ 
for the sake of brevity. The 4D Levi-Civita tensor $\varepsilon_{\mu\nu\eta\kappa}$ 
(with $\mu, \nu, \eta...= 0, 1, 2, 3$) satisfies
$\varepsilon_{\mu\nu\eta\kappa} \varepsilon^{\mu\nu\eta\kappa} = - 4!, 
\varepsilon_{\mu\nu\eta\kappa} \varepsilon^{\mu\nu\eta\sigma} = - 3! \delta^\sigma_\kappa$, 
etc., and $\varepsilon_{0123} = +1$. We shall be using the above convention of dot and cross products
throughout the whole body of our text in the description of our 
present SU(N) gauge theory.\\

\begin{center}
{\section{Preliminaries: gauge symmetry transformations}}
\end{center}

We begin with the following basic Lagrangian density for the 4D
topologically massive non-Abelian gauge theory with mass parameter $m$ (see, e.g.  [7,8] for details)
\begin{eqnarray}
{\cal L}_{(0)} = - \frac{1}{4}\; F^{\mu\nu} \cdot F_{\mu\nu} + \frac{1}{12} \;H^{\mu\nu\eta} \cdot H_{\mu\nu\eta}
+ \frac{m}{4} \;\varepsilon_{\mu\nu\eta\kappa} B^{\mu\nu} \cdot F^{\eta\kappa},
\end{eqnarray}
where the 2-form $F^{(2)} = d A^{(1)} + i \;A^{(1)} \wedge A^{(1)} \equiv 
\frac{1}{2!} (dx^\mu \wedge dx^\nu) F_{\mu\nu} \cdot T$ defines the curvature tensor
$F_{\mu\nu} = \partial_\mu A_\nu - \partial_\nu A_\mu - (A_\mu \times A_\nu)$ for the 1-form
($A^{(1)} = dx^\mu A_\mu \cdot T$) gauge potential $A_\mu$ and 3-form $H^{(3)} = \frac{1}{3!}\;
(dx^\mu \wedge dx^\nu \wedge dx^\eta)\; H_{\mu\nu\eta} \cdot T$ defines the compensated curvature tensor
in terms of the dynamical 2-form ($B^{(2)} = \frac{1}{2!} (dx^\mu \wedge dx^\nu)\; B_{\mu\nu}$) gauge potential
$B_{\mu\nu}$ and the 1-form ($K^{(1)} =  dx^\mu \;K_\mu \cdot T$) compensating auxiliary field $K_\mu$,
as
\begin{eqnarray}
H^a_{\mu\nu\eta} &=& (\partial_\mu B^a_{\nu\eta} + \partial_\nu B^a_{\eta\mu} + \partial_\eta B^a_{\mu\nu})
- \bigl [(A_\mu \times B_{\nu\eta})^a + (A_\nu \times B_{\eta\mu})^a + (A_\eta \times B_{\mu\nu})^a \bigr ] \nonumber\\
&-& \bigl [(K_\mu \times F_{\nu\eta})^a + (K_\nu \times F_{\eta\mu})^a + (K_\eta \times F_{\mu\nu})^a \bigr ].
\end{eqnarray}
The last term in the above Lagrangian  density ${\cal L}_{(0)}$
corresponds to the topological mass term where the curvature
tensor $F_{\mu\nu}$ (corresponding to the 
non-Abelian 1-form gauge field) and the dynamical 2-form gauge field $B_{\mu\nu}$
are coupled together through $B^{(2)}\wedge F^{(2)}$ term.

The above starting Lagrangian density ${\cal L}_{(0)}$
respects the {\it usual} infinitesimal ``scalar''
gauge transformations $\delta_g $ corresponding to the non-Abelian 1-form gauge theory. In fact, under this transformation, the relevant 
fields of (1) (and (1) itself) transform as [13]
\begin{eqnarray}
&\delta_g A_\mu = D_\mu \Omega \equiv \partial_\mu \Omega - (A_\mu \times \Omega),\qquad
\delta_g F_{\mu\nu} = - (F_{\mu\nu} \times \Omega), \qquad  
\delta_g B_{\mu\nu} = - (B_{\mu\nu} \times \Omega), & \nonumber\\
& \delta_g H_{\mu\nu\eta} = - (H_{\mu\nu\eta} \times \Omega), \qquad \delta_g K_\mu = - (K_\mu \times \Omega),
\qquad \delta_g {\cal L}_{(0)} = 0,
\end{eqnarray}
where $\Omega = \Omega \cdot T$ is the infinitesimal
$SU(N)$-valued ``scalar'' gauge parameter. It is evident, from the above equation, that the basic Lagrangian density (1)
remains invariant under $\delta_g $.

In addition to (3), there exists an
independent vector gauge symmetry transformation $\delta_v$
(parametrized by a 4D vector infinitesimal gauge parameter $\Lambda_\mu = \Lambda_\mu \cdot T$),
under which, the fields of the above Lagrangian density ${\cal L}_{(0)}$ transform as (see, e.g. [13])
\begin{eqnarray}
\delta_v A_\mu = 0, \quad \delta_v K_\mu = - \Lambda_\mu\ , 
\quad \delta_v B_{\mu\nu} = -(D_\mu \Lambda_\nu - D_\nu \Lambda_\mu ), 
\quad\delta_v F_{\mu \nu}=0,\quad \delta_v H_{\mu \nu \eta}=0 .
\end{eqnarray}
It is straightforward to check that the Lagrangian density (1) transforms to a total spacetime derivative under (4).
This can be mathematically expressed as
\begin{eqnarray}
\delta_v {\cal L}_{(0)} = - m\; \partial_\mu\; \Bigl [ \;\varepsilon^{\mu\nu\eta\kappa} 
\Lambda_\nu \cdot \Bigl (  \partial_\eta A_\kappa - \frac{1}{2}\; A_\eta \times A_\kappa
\Bigr ) \; \Bigr ].
\end{eqnarray}
Thus, the action integral of the present theory would remain invariant under the above
local, infinitesimal ``vector'' gauge symmetry transformations. \\


\begin{center}
{\section {``Scalar'' gauge symmetry transformations and superfield formalism: a brief sketch}}
\end{center}

For the paper to be self-contained, we discuss here the bare essentials of the key ideas
that have been exploited in our earlier work [18]. It is to be noted that, under the
``scalar'' gauge symmetry transformations (3), the kinetic term corresponding to the
1-form gauge field remains invariant [i.e. $\delta_g \;(F^{\mu\nu} \cdot F_{\mu\nu}) = 0$]. 
As a consequence, when we generalize the 4D
ordinary theory onto the (4, 2)-dimensional supermanifold, we invoke the horizontality
condition that, ultimately, implies the following gauge invariant  
equality [14]
\begin{eqnarray}
- \;\frac{1}{4}\; \tilde {\cal F}^{MN} (x,\theta,\bar\theta) 
\cdot \tilde {\cal F}_{MN} (x,\theta,\bar\theta)
= - \;\frac{1}{4} \;F^{\mu\nu} \cdot F_{\mu\nu},
\end{eqnarray}
where the super-curvature tensor $\tilde {\cal F}_{MN}$ on the supermanifold is defined through the super 2-form $
\tilde {\cal F}^{(2)} = \frac{1}{2!}\; (dZ^M \wedge dZ^N)\; \tilde{\cal F}_{MN} \equiv
\tilde d \tilde {\cal A}^{(1)} + i\; \tilde {\cal A}^{(1)} \wedge \tilde{\cal A}^{(1)}$.
The above (4, 2)-dimensional supermanifold is parametrized by the superspace variables
$Z^M = (x^\mu, \theta, \bar\theta)$ and the nilpotent ($\tilde d^2 = 0$)
super exterior derivative $\tilde d$ and super 1-form connection $\tilde {\cal A}^{(1)}$
are defined as [14]
\begin{eqnarray}
\tilde d = d Z^M \partial_M &\equiv& dx^\mu \;\partial_\mu + d \theta \;\partial_\theta
+ d \bar \theta \;\partial_{\bar\theta},\; \qquad 
\partial_M = (\partial_\mu, \partial_\theta, \partial_{\bar\theta}), \nonumber\\
\tilde {\cal A}^{(1)} = d Z^M A_M &\equiv& dx^\mu \;\tilde {\cal B}_\mu (x,\theta,\bar\theta) 
+ d \theta \;  {\tilde {\bar{\cal F}}} (x,\theta,\bar\theta) 
+ {d \bar\theta}\; {\tilde {\cal F}} (x,\theta,\bar\theta ), 
\end{eqnarray} 
where the superfields $\tilde {\cal B}_\mu (x,\theta,\bar\theta)$, $\tilde {\cal F} (x,\theta,\bar\theta)$
and ${\tilde {\bar {\cal F}}}(x,\theta,\bar\theta)$ are the generalization of the  1-form gauge field
$A_\mu (x)$, ghost field $C (x)$ and anti-ghost field $\bar C (x)$ of the BRST invariant ordinary 4D non-Abelian
gauge theory [21] onto the (4, 2)-dimensional supermanifold.

The above statement is corroborated by the following expansions of the superfields
along the Grassmannian directions $\theta$ and $\bar\theta$ of the supermanifold  (see, e.g. [14])
\begin{eqnarray}
\tilde {\cal B}_\mu (x, \theta, \bar \theta) &=& A_\mu (x) + \theta \;\bar R_\mu (x) + \bar \theta \;R_\mu (x)
+ i \;\theta \;\bar\theta \;S_\mu (x) \nonumber\\
& \equiv & A_\mu (x) + \theta\; \Bigl (s_{ab}^{(1)} A_\mu (x) \Bigr ) 
+ \bar\theta\; \Bigl (s_b^{(1)} A_\mu (x) \Bigr )
+ \theta\;\bar\theta\; \Bigl (s_b^{(1)} s_{ab}^{(1)} A_\mu (x)\Bigr ),\; \nonumber\\
\tilde {\cal F} (x, \theta, \bar \theta) &=& C (x) + i \;\theta \; \bar B_1 (x) + i \;\bar\theta \;B_1 (x)
+ i \;\theta \;\bar\theta\; s(x) \nonumber\\
& \equiv & C (x) + \theta\; \Bigl (s_{ab}^{(1)} C (x) \Bigr ) + \bar\theta\; \Bigl (s_b^{(1)} C (x) \Bigr )
+ \theta\;\bar\theta\; \Bigl (s_b^{(1)} s_{ab}^{(1)} C (x) \Bigr ), \nonumber\\
{\tilde {\bar {\cal F}}} (x, \theta, \bar \theta) &=& \bar C (x) + i \;\theta \;\bar B_2 (x) + i\; \bar\theta \;B_2 (x)
+ i \;\theta \;\bar\theta \;\bar s (x) \nonumber\\
& \equiv & \bar C (x) + \theta\; \Bigl (s_{ab}^{(1)} \bar C (x) \Bigr ) 
+ \bar\theta\; \Bigl (s_b^{(1)} \bar C (x) \Bigr )
+ \theta\;\bar\theta\; \Bigl (s_b^{(1)} s_{ab}^{(1)} \bar C (x) \Bigr ), 
\end{eqnarray}
where the secondary fields ($\bar R_\mu (x), R_\mu (x), s(x), \bar s(x)$) are fermionic and the
other secondary fields ($S_\mu (x), B_1 (x), \bar B_1 (x), B_2 (x), \bar B_2 (x)$) are bosonic in 
nature.
These secondary fields are determined in terms of the basic and auxiliary fields of the ordinary 4D non-Abelian
1-form gauge theory by exploiting the mathematical power of horizontality condition (HC). It should be noted that, 
in equation (8), we have identified the (anti-)BRST symmetry transformations, corresponding to the
usual ``scalar" gauge symmetry transformations, by the standard notation $s_{(a)b}^{(1)}$ (see e.g. [18-20]). In fact,
the HC (i.e. $\tilde {\cal F}^{(2)} = F^{(2)}$) leads to the derivation of the following relationships [14]
\begin{eqnarray}
&& R_\mu = D_\mu C,\; \quad \bar R_\mu = D_\mu \bar C,\; \quad B_1 = - \frac{i}{2}\; (C \times C),\;
\quad  s = - (\bar B_1 \times C), \nonumber\\
&& S_\mu = D_\mu B_2 + i\; (D_\mu C \times \bar C) \equiv - D_\mu \bar B_1 - i\; (C \times D_\mu \bar C), \nonumber\\
&& \bar B_2 = - \frac{i}{2}\; (\bar C \times \bar C),\; \quad
\bar B_1 + B_2 = - i\; (C \times \bar C),\; \quad \bar s = - (B_2 \times \bar C).
\end{eqnarray}
If we make the identifications: $ \bar B_1 = \bar B, B_2 = B$, the above Curci-Ferrari (CF) restriction 
$\bar B_1 + B_2 = - i \;(C \times \bar C)$ changes over to $B + \bar B = - i \;(C \times \bar C)$. This condition
plays an important role in the proof of the {\it absolute} anticommutativity of the (anti-)BRST
symmetry transformations for the 4D non-Abelian 1-form gauge theory (e.g. $\{ s_b^{(1)}, s_{ab}^{(1)} \} A_\mu =0$
only when $B + \bar B = - i \;(C \times \bar C)$). Furthermore, the above CF condition leads to the
existence of coupled Lagrangian densities for the theory.

The insertions of the above values of secondary fields in (8) leads to the derivation
of the off-shell nilpotent and absolutely anticommuting (anti-)BRST symmetry transformations
\begin{eqnarray}
&& s_b^{(1)} A_\mu = D_\mu C,\; \qquad s_b^{(1)} C = \frac{1}{2} (C \times C),\; \qquad
s_b^{(1)} \bar C = i\; B,\;  \qquad s_b^{(1)} B = 0, \nonumber\\ &&s_b^{(1)} \bar B = - (\bar B \times C),\; 
\qquad s_{ab}^{(1)} A_\mu = D_\mu \bar C,\; \qquad s_{ab}^{(1)} \bar C = \frac{1}{2} (\bar C \times \bar C), \nonumber\\
&& s_{ab}^{(1)} C = i\; \bar B,\; \qquad s_{ab}^{(1)} \bar B = 0,\; \qquad s_{ab}^{(1)} B = - (B \times \bar C), 
\end{eqnarray}
because the following mapping is valid between the translational generators 
($\partial_\theta, \partial_{\bar\theta}$) along the Grassmannian
directions and the (anti-)BRST symmetry transformations, namely;
\begin{eqnarray}
\lim_{\theta \to 0} \; \frac{\partial}{\partial \bar\theta}\; \tilde \Sigma^{(h)} (x, \theta, \bar\theta)
= s_b^{(1)} \;\Sigma (x),\; \qquad \qquad 
\lim_{\bar \theta \to 0} \; \frac{\partial}{\partial \theta}\; \tilde \Sigma^{(h)} (x, \theta, \bar\theta)
= s_{ab}^{(1)} \;\Sigma (x),
\end{eqnarray}
where the superscript ${(h)}$ on the generic superfield $\tilde \Sigma (x,\theta,\bar\theta)$ 
denotes the corresponding 
superfield obtained after the application of the HC and the 4D generic local field $\Sigma (x)$
corresponds to the local fields of the 4D ordinary non-Abelian 1-form gauge theory. The (anti-)BRST
symmetry transformations on the auxiliary fields $B$ and $\bar B$ in (10) have been derived due
to the requirement of absolute anticommutativity between $s_b^{(1)}$ and $s_{ab}^{(1)}$.

The spacetime component
of the super-curvature tensor $\tilde {\cal F}_{MN}$, after the application of the HC,
 is $\tilde {\cal F}^{(h)}_{\mu\nu} (x,\theta,\bar\theta)
= \partial_\mu \tilde {\cal B}^{(h)}_\nu - \partial_\nu \tilde {\cal B}^{(h)}_\mu 
+ i [ \tilde {\cal B}^{(h)}_\mu, \tilde {\cal B}^{(h)}_\nu ]$. This can be written, using the expansion
for $\tilde {\cal B}^{(h)}_\mu (x,\theta,\bar\theta)$ (from equation (8)) as follows:
\begin{eqnarray}
\tilde {\cal F}^{(h)}_{\mu\nu} (x,\theta,\bar\theta) &=& F_{\mu\nu} - \theta \;(F_{\mu\nu} \times \bar C)
- \bar \theta \;(F_{\mu\nu} \times  C) + \theta\;\bar \theta \;[ (F_{\mu\nu} \times C) \times \bar C - i\; 
F_{\mu\nu} \times B] \nonumber\\
&\equiv& F_{\mu\nu} (x) + \theta \; \Bigl ( s_{ab}^{(1)} F_{\mu\nu} (x) \Bigr ) + \bar \theta\; \Bigl (s_b^{(1)} F_{\mu\nu} (x) \Bigr )
+ \theta \;\bar \theta\; \Bigl (s_b^{(1)} s_{ab}^{(1)} F_{\mu\nu} (x) \Bigr ).
\end{eqnarray}
The above expression implies clearly that the kinetic term remains invariant under the horizontality
condition ($- \frac{1}{4} \tilde {\cal F}^{\mu\nu (h)} (x,\theta,\bar\theta) 
\cdot \tilde{\cal F}^{(h)}_{\mu\nu} (x,\theta,\bar\theta)
= - \frac{1}{4} F^{\mu\nu} \cdot F_{\mu\nu}$). Furthermore, the above expansion yields the (anti-)BRST 
transformations  for the curvature tensor $F_{\mu\nu}$ as
\begin{eqnarray}
&&s_b^{(1)} F_{\mu\nu} = - (F_{\mu\nu} \times C),\; \qquad s_{ab}^{(1)} F_{\mu\nu} = - (F_{\mu\nu} \times \bar C),
\nonumber\\
&& s_b^{(1)} s_{ab}^{(1)} F_{\mu\nu} = (F_{\mu\nu} \times C) \times \bar C - i\; 
F_{\mu\nu} \times B. 
\end{eqnarray}
It is interesting to note that we have the gauge-invariant quantities that incorporate the curvature tensor $F_{\mu\nu}$. For instance, it can be checked that
$ \delta_g \;(B_{\mu\nu} \cdot F_{\eta\kappa}) = 0, 
\delta_g \; (K_{\mu} \cdot F_{\nu\eta}) = 0$ under the transformations (3). As a
consequence, these remain invariant when we generalize the 4D local fields onto the (4, 2)-dimensional
supermanifold with the corresponding superfields. 
Thus, we propose the following gauge-invariant restrictions (GIRs) [18]
\begin{eqnarray}
 \tilde {\cal B}_{\mu\nu} (x,\theta,\bar\theta) \cdot \tilde {\cal F}^{(h)}_{\eta\kappa} (x,\theta,\bar\theta)
&=& B_{\mu\nu} (x) \cdot F_{\eta\kappa} (x), \nonumber\\
\tilde {\cal K}_{\mu} (x,\theta,\bar\theta) \cdot \tilde {\cal F}^{(h)}_{\nu\eta} (x,\theta,\bar\theta)
&=& K_{\mu} (x) \cdot F_{\nu\eta} (x),
\end{eqnarray}
as the analogues of the gauge invariant horizontality restriction (6).

We take the  expansions of the superfields $\tilde {\cal B}_{\mu\nu} (x,\theta,\bar\theta)$ and
$\tilde {\cal K}_{\mu} (x,\theta,\bar\theta)$ along the Grassmannian
directions of the (4, 2)-dimensional supermanifold as
\begin{eqnarray}
\tilde {\cal B}_{\mu\nu} (x, \theta, \bar \theta) &=& B_{\mu\nu} (x) + \theta \; \bar R_{\mu\nu} (x) + \bar \theta \;
R_{\mu\nu} (x)
+ i\;\theta \;\bar\theta \; S_{\mu\nu} (x), \nonumber\\
\tilde {\cal K}_\mu (x, \theta, \bar \theta) &=& K_\mu (x) + \theta \; \bar P_\mu (x) + \bar\theta \; 
P_\mu (x) + i\;\theta \;\bar\theta\;Q_\mu (x), 
\end{eqnarray}
where the secondary fields ($R_{\mu\nu}, \bar R_{\mu\nu}, P_\mu, \bar P_\mu$) are fermionic 
and ($S_{\mu\nu}, Q_\mu$) are bosonic in nature. It is straightforward to check that the 
following relationships ensue from (14):
\begin{eqnarray}
&& R_{\mu\nu} = - (B_{\mu\nu} \times C),\; \quad  \bar R_{\mu\nu} = - (B_{\mu\nu} \times \bar C),\; \quad
S_{\mu\nu} = - (B_{\mu\nu} \times B) - i \; \Bigl [(B_{\mu\nu} \times C) \times \bar C \Bigr ], \nonumber\\
&& P_\mu  = - (K_\mu \times C),\; \quad \bar P_\mu = - (K_\mu \times \bar C),\; \quad
Q_\mu = - (K_{\mu} \times B) - i \;\Bigl [ (K_{\mu} \times C) \times \bar C\Bigr ].
\end{eqnarray}
Taking the help of mapping in (11), it is clear that we have the following off-shell nilpotent 
and anticommuting (anti-)BRST transformations for the relevant fields
\begin{eqnarray}
&& s_b^{(1)} H_{\mu\nu\eta} = - (H_{\mu\nu\eta} \times C),\; \qquad
s_b^{(1)} B_{\mu\nu} = - (B_{\mu\nu} \times C),\; \qquad s_b^{(1)} K_\mu = - (K_\mu \times C), \nonumber\\
&& s_{ab}^{(1)} H_{\mu\nu\eta} = - (H_{\mu\nu\eta} \times \bar C),\; \quad
s_{ab}^{(1)} B_{\mu\nu} = - (B_{\mu\nu} \times \bar C),\; \quad s_{ab}^{(1)} K_\mu = - (K_\mu \times \bar C),
\end{eqnarray}
where the transformations for the curvature tensor $H_{\mu\nu\eta}$ have been derived
from the following expansion (where the expansions of constituent fields have been taken into account):
\begin{eqnarray}
&&\tilde {\cal H}^{(g,h)}_{\mu\nu\eta}\; (x, \theta, \bar \theta) = H_{\mu\nu\eta} (x) - \;\theta \; 
[(H_{\mu\nu\eta} \times \bar C) (x)] - \;\bar \theta \; [(H_{\mu\nu\eta} \times  C) (x)] \nonumber\\
&& + \;  \theta \;\bar\theta \; \Bigl [ \Bigl \{(H_{\mu\nu\eta} \times C) \times \bar C 
- i \; H_{\mu\nu\eta} \times B \Bigr \} (x) \Bigr ]
\nonumber\\
&& \equiv  H_{\mu\nu\eta} (x) + \theta\; \Bigl (s_{ab}^{(1)} H_{\mu\nu\eta} (x)\Bigr ) 
+ \bar\theta\; \Bigl( s_b^{(1)} H_{\mu\nu\eta} (x) \Bigr)
+ \theta\;\bar\theta\; \Bigl (s_b^{(1)} s_{ab}^{(1)} H_{\mu\nu\eta} (x) \Bigr ).
\end{eqnarray}
Here the superscripts ${(g, h)}$ denote the expansion derived after the application of the GIRs and HC.
Thus, we have derived {\it all} the (anti-)BRST transformations for $\it all$ the fields.

Before we wrap up this section, we would like to lay emphasis on the equations (10), (13) and (17)
which encapsulate all the off-shell nilpotent and absolutely anticommuting (anti-) BRST symmetry
transformations corresponding to the ``scalar'' gauge transformations (3) of the theory. A key
point, in our whole discussion, is worth noting. We are theoretically compelled to go beyond the
application of the HC and we are forced to invoke some GIRs (cf. (14)) in order to obtain the whole set of
(anti-)BRST symmetry transformations in the context of a gauge theory (where there is no interaction with
matter fields). This is a {\it new} observation in the context of superfield formulation of a 
gauge theory without matter fields. In our earlier work [17], the coupled Lagrangian densities have been 
derived that respect the (anti-)BRST symmetry transformations listed in (10), (13) and (17) {\it together}. As a consequence, 
we do not  elaborate on these issues in our present section.\\

\begin{center}
{\section {``Vector'' gauge symmetry transformations and superfield formalism: a detailed discussion}}
\end{center}

It is evident from equation (4), corresponding to the``vector'' gauge symmetry transformations, that the field $A_\mu$,
curvature tensors $F_{\mu \nu}$ and $H_{\mu \nu \eta}$ are gauge invariant quantities (i.e. $\delta_v A_{\mu}=0,
\;\delta_v F_{\mu \nu}=0 ,\;\delta_v H_{\mu \nu \eta}=0)$ . As a consequence, they remain {\it invariant} when we generalise 
the 4D ordinary theory onto a (4, 2)-dimensional supermanifold. Mathematically, the above statement of gauge invariant restrictions (GIRs)
can be expressed as:
\begin{eqnarray}
\tilde {\cal A}^{(1)} =  A^{(1)},\; \qquad \tilde{\cal F}^{(2)} =  F^{(2)},\; \qquad \tilde {\cal H}^{(3)} =  H^{(3)}.
\end{eqnarray}
As a result of the above equality, we have the following relationships from the first two gauge invariant restrictions (cf. expansions in (8))
\begin{eqnarray}
&& \tilde {\cal B}_{\mu}^{(g)}(x,\theta,\bar \theta) = A_{\mu}(x),
\;  \qquad \tilde {\cal F}^{(g)}(x,\theta,\bar \theta) = 0,\; \qquad 
{\tilde {\bar{\cal F}}}^{(g)}(x,\theta,\bar \theta) =  0, \nonumber\\ 
&& \partial_{\mu} \tilde {\cal B}_{\nu}^{(g)}-\partial_{\nu}\tilde {\cal B}_\mu^{(g)} 
+ i \; [\tilde {\cal B}_\mu^{(g)}, \; \tilde {\cal B}_\nu^{(g)} ]= F_{\mu \nu}(x),
\end{eqnarray}
where the superscript {\it(g)} on superfields corresponds to the restrictions that have been obtained after 
the applications of GIR (cf. 19). The l.h.s. and r.h.s. of the third GIR of (19) can be expressed as follows:
\begin{eqnarray}
H^{(3)} &=& dB^{(2)} + i\;(A^{(1)} \wedge B^{(2)} - B^{(2)} \wedge A^{(1)}) + i\;( K^{(1)} \wedge F^{(2)} 
- F^{(2)} \wedge  K^{(1)}), \nonumber\\
\tilde {\cal H}^{(3)} &=& \tilde {d} \tilde {\cal B}^{(2)}+ i\;(A^{(1)} \wedge \tilde {\cal B}^{(2)} 
- \tilde {\cal B}^{(2)} \wedge A^{(1)})
+i \;(\tilde {\cal K}^{(1)} \wedge F^{(2)} - F^{(2)} \wedge \tilde {\cal K}^{(1)}),
\end{eqnarray}
where it is elementary to check that $H^{(3)}$ produces the curvature tensor (2). In the above, we have 
taken $\tilde {\cal A}^{(1)}\;= \;A^{(1)}$ and $\tilde {\cal F}^{(2)}\;=\; F^{(2)}$ (cf. 19) and other expansions are [22]
\begin{eqnarray}
\tilde {\cal K}^{(1)}&=& dx^{\mu}\;\tilde {\cal K}_{\mu}(x, \theta, \bar\theta) 
+ d\theta \;{\tilde {\bar{\cal F}}}_{1}(x, \theta, \bar\theta) + {d\bar\theta} \;\tilde {\cal F}_{1}(x,\theta,\bar\theta), \nonumber\\
\tilde {\cal B}^{(2)}&=&\frac{1}{2!}\; (dZ^M \wedge dZ^N)\;\tilde {\cal B}_{MN}(x, \theta,\bar\theta)\nonumber\\
&\equiv & \frac{1}{2!} \;(dx^{\mu}\wedge dx^{\nu})\; \tilde {\cal B}_{\mu \nu}(x, \theta,\bar\theta) 
+(dx^{\mu} \wedge d\theta)\; {\tilde {\bar{\cal F}}}_{\mu}(x,\theta,\bar\theta)\nonumber\\
&+&(dx^\mu \wedge d\bar\theta) \;\tilde {\cal F}_\mu (x,\theta,\bar\theta)
+(d\theta \wedge d\bar\theta)\;\tilde\Phi(x,\theta,\bar\theta)\nonumber\\
&+&(d\theta \wedge d\theta)\;{\tilde{\bar{\cal \beta}}}(x,\theta,\bar\theta) 
+ (d\bar\theta \wedge d\bar\theta)\;\tilde {\cal \beta}(x,\theta,\bar\theta).
\end{eqnarray}
The above multiplet superfields in the expansion of $\tilde {\cal B}^{(2)}$ and $\tilde {\cal K}^{(1)}$ 
are to be expanded along the Grassmannian directions $\theta$ and $\bar\theta$
in terms of the basic fields $B_{\mu \nu}, \; \bar C_\mu,\; C_\mu,\; \phi,\; \beta,\; \bar\beta$
of the BRST invariant non-Abelian 2-form gauge theory [22] and (anti-)ghost fields $\bar C_1$ and $C_1$
corresponding to the compensating  auxiliary field $K_\mu$.
Consistent with our earlier work [22], we have the following expansions for the superfields in addition to (15)
and (22):
\begin{eqnarray}
\tilde {\cal F}_{\mu}(x, \theta, \bar\theta) &=& C_\mu (x)+ \theta \;{\bar B}^{(1)}_\mu(x) 
+ \bar \theta \; B^{(1)}_\mu(x) +i\; \theta\;\bar\theta \;S_\mu(x), \nonumber\\
{\tilde {\bar{\cal F}}}_{\mu}(x, \theta, \bar\theta) &=& {\bar C}_\mu (x)+ \theta \; \bar B^{(2)}_\mu(x) 
+ \bar \theta \; B^{(2)}_\mu(x) +i\; \theta\;\bar\theta \; {\bar S}_\mu(x),  \nonumber\\
\tilde \Phi(x, \theta, \bar\theta)&=& \phi(x) + \theta \; {\bar f}_1(x) + \bar\theta \; f_1(x) 
+ i\; \theta\;\bar\theta \; b_1(x), \nonumber \\
\tilde \beta (x, \theta, \bar\theta)&=& \beta(x) + \theta \;{\bar f}_2(x) + \bar\theta \; f_2(x) 
+ i\; \theta\;\bar\theta \; b_2(x), \nonumber\\
\tilde {\bar \beta} (x, \theta, \bar\theta)&=& {\bar \beta}(x) + \theta \; {\bar f}_3(x) + \bar\theta \; f_3(x) 
+ i\; \theta\;\bar\theta \; b_3(x), \nonumber\\
\tilde {\cal F}_1 (x, \theta, \bar\theta)&=& C_1(x) + i\;\theta \; {\bar R}(x) + i\;\bar\theta \; R(x) 
+ i\; \theta\;\bar\theta \; s_1(x), \nonumber\\
{\tilde {\bar{\cal F}}}_1 (x, \theta, \bar\theta)&=& {\bar C_1}(x) + i\;\theta \; {\bar S}(x) + i\;\bar\theta \; S(x) 
+ i\; \theta\;\bar\theta \;{\bar s}_1(x),
\end{eqnarray}
where  all the secondary fields on the r.h.s. of the above super-expansion would be determined in terms of the basic and auxiliary 
fields of the 4D ordinary theory by exploiting the horizontality-type requirement ($ \tilde {\cal H}^{(3)} =  H^{(3)}$). The secondary fields 
$( S_\mu, {\bar S}_\mu, f_1, {\bar f}_1, f_2, {\bar f}_2, f_3, {\bar f}_3, s_1, {\bar s}_1 )$ and
$( B_\mu^{(1)}, {\bar B}_\mu^{(1)}, B_\mu^{(2)}, {\bar B}_\mu^{(2)}, b_1, b_2, b_3, R, \bar R, S, \bar S )$ are fermionic and bosonic, respectively.

To obtain the explicit expressions for the secondary fields in terms of the basic fields and auxiliary fields, 
we have to express each term of $\tilde {\cal H}^{(3)}$ (cf. 21). The first term is 
\begin{eqnarray}
&&\tilde d \tilde {\cal B}^{(2)} = \frac {1}{3!}\; (dx^ {\mu} \wedge dx ^{\nu} \wedge dx^{\eta})\;(\partial_{\mu} \tilde {\cal B}_{\nu \eta}
+ \partial_{\nu} \tilde {\cal B}_{\eta\mu} + \partial_{\eta} \tilde {\cal B}_{\mu \nu}) 
+ (d \theta \wedge d \theta \wedge d \theta )\; (\partial_{\theta} \tilde {\bar \beta}) \nonumber\\ 
&& + (d \bar\theta \wedge d \bar\theta \wedge d \bar\theta )\; (\partial_{\bar\theta} \tilde \beta )
+ (d \theta \wedge d \bar\theta \wedge d \bar\theta ) \;(\partial_{\theta} \tilde \beta + \partial_{\bar\theta} \tilde \Phi )
+ (d \bar\theta \wedge d \theta \wedge d \theta )\; (\partial_{\bar\theta} \tilde {\bar \beta}
+ \partial_{\theta} \tilde \Phi )\nonumber\\
&&+ (dx^{\mu} \wedge d \bar\theta \wedge d \bar\theta )\;(\partial_{\mu} \tilde\beta 
+ \partial_{\bar\theta} \tilde {\cal F}_{\mu}) + \frac {1}{2!}\;(dx^{\mu} \wedge dx^{\nu} \wedge d\theta )\;(\partial_{\mu} \tilde {\bar{\cal  F}}_{\nu}} 
- \partial_{\nu} \tilde {\bar{\cal F}}_{\mu} + \partial_{\theta} {\tilde{\cal B}_{\mu \nu}) \nonumber\\
&&+ (dx^{\mu} \wedge d\theta \wedge d\bar\theta )\;(\partial_{\mu} \tilde\Phi + \partial_{\theta} \tilde {\cal F}_{\mu} 
+ \partial_{\bar\theta} {\tilde {\bar {\cal F}}}_{\mu}) + (dx^{\mu} \wedge d\theta \wedge d\theta )\;(\partial_{\mu} \tilde{\bar\beta} 
+ \partial_{\theta} {\tilde {\bar {\cal F}}}_\mu )\nonumber\\
&&+ \frac {1}{2!}\; (dx^{\mu} \wedge dx^{\nu} \wedge d\bar\theta)\;(\partial_\mu \tilde {\cal F}_\nu - \partial_\nu \tilde {\cal F}_\mu
+ \partial_{\bar \theta} \tilde {\cal B}_{\mu \nu}).
\end{eqnarray}
The second term of the super 3-form $\tilde {\cal H}^{(3)}$ is given by
\begin{eqnarray}
&&i\;[( A^{(1)} \wedge \tilde {\cal B}^{(2)} ) - ( \tilde {\cal B}^{(2)} \wedge A^{(1)})] = -\;\frac{1}{3!} \;(dx^{\mu} \wedge dx^{\nu} \wedge dx^{\eta})\;
[(A_\mu \times \tilde {\cal B}_{\nu \eta} )\nonumber\\ 
&&+ (A_\nu \times \tilde{\cal B}_{\eta \mu} ) + (A_\eta \times \tilde{\cal B}_{\mu \nu} )]
- \;\frac{1}{2!}\; (dx^{\mu} \wedge dx^{\nu} \wedge d \theta )\; [(A_\mu \times \tilde {\bar{\cal F}}_\nu ) 
- (A_\nu \times \tilde {\bar{\cal F}}_\mu )]\nonumber\\
&&- \;\frac{1}{2!}\; (dx^{\mu} \wedge dx^{\nu} \wedge d \bar\theta ) \;[(A_\mu \times \tilde {{\cal F}}_\nu ) 
- (A_\nu \times \tilde {\cal F}_\mu ) ] -(dx^{\mu} \wedge d\theta \wedge d\bar\theta )\;(A_\mu \times \tilde \Phi )\nonumber\\
&&- (dx^{\mu } \wedge d \theta \wedge d \theta ) \;(A_\mu \times \tilde{\bar \beta})
- (dx^{\mu } \wedge d \bar\theta \wedge d \bar\theta ) \;(A_\mu \times {\tilde \beta}).
\end{eqnarray}
The last term of the $ \tilde {\cal H}^{(3)}$ is as follows
\begin{eqnarray}
&& i[(\tilde {\cal K}^{(1)} \wedge F^{(2)} ) - ( F^{(2)} \wedge \tilde {\cal K}^{(1)})] 
= -\frac{1}{3!} \;(dx^{\mu} \wedge dx^{\nu} \wedge dx^{\eta})\;
[(\tilde {\cal K}_\mu \times  F_{\nu \eta} )\nonumber\\ 
&& + (\tilde {\cal K}_\nu \times F_{\eta \mu} ) + (\tilde {\cal K}_\eta \times  F_{\mu \nu} )] 
-\frac{1}{2!} \;(dx^\mu \wedge dx^\nu \wedge d\theta )\;({\tilde {\bar{\cal F}}}_1 \times F_{\mu \nu})\nonumber\\
&& -\frac{1}{2!} \;(dx^\mu \wedge dx^\nu \wedge d \bar\theta )\;(\tilde {\cal F}_1 \times F_{\mu \nu}). 
\end{eqnarray}
In the requirement of HC, the sum of (24), (25) and (26) produces the super 3-form $\tilde {\cal H}^{(3)}$.
The key point is the fact that gauge invariant curvature tensor $ H_{\mu\nu\eta}$
remains unaffected due to the presence of the 
Grassmannian variables $\theta$ and $\bar\theta$. As a result, we set equal to zero the 
coefficients of all the differentials of the super 3-form that incorporate the differentials of 
the Grassmannian variables (i.e. $d\theta$ 
and $d\bar\theta$). To achieve  this goal and to pin-point all the differentials that incorporate the Grassmannian variables, 
we have to have a close look at all the differentials on the r.h.s. of (24),
(25) and (26). Mathematically, it means that we have to set the coefficient of differentials 
$(dx^{\mu} \wedge dx^{\nu} \wedge d\bar\theta ), (dx^{\mu} \wedge dx^{\nu} \wedge d \theta ),
(dx^{\mu} \wedge d\theta \wedge d\bar \theta ),(dx^{\mu} \wedge d\bar\theta \wedge d\bar \theta ),
(d\theta \wedge d\theta \wedge d \theta ),(dx^{\mu} \wedge d\theta \wedge d\theta ),  
(d\theta \wedge d\theta \wedge d\bar\theta ), (d\theta \wedge d\bar\theta \wedge d \bar\theta ),
(d\bar\theta \wedge d\bar\theta \wedge d \bar\theta )$ equal to zero.
These leads to the following $\it nine$ relationships amongst the multiplet superfields
\begin{eqnarray}
& D_\mu \tilde {\cal F}_\nu - D_\nu \tilde {\cal F}_\mu + \partial_{\bar\theta} \tilde {\cal B}_{\mu\nu} 
+ i\; \Bigl [\tilde {\cal F}_{1}, F_{\mu\nu}\Bigr ] = 0, & \nonumber\\
& D_\mu {\tilde {\bar{\cal F}}}_\nu - D_\nu {\tilde {\bar{\cal F}}}_\mu + \partial_{\theta} {\tilde {\cal B}}_{\mu\nu} 
+ i\; \Bigl[{\tilde {\bar{\cal F}}}_{1}, F_{\mu\nu} \Bigr ] = 0, & \nonumber\\
& D_\mu {\tilde {\Phi}} + \partial_{\bar\theta} {\tilde {\bar{\cal F}}}_{\mu} + \partial_{\theta} 
\tilde {\cal F}_\mu = 0, \; \qquad
 D_\mu \tilde \beta + \partial_{\bar\theta} \tilde {\cal F}_\mu = 0,
\; \qquad \partial_\theta \tilde {\bar\beta}  = 0, &\nonumber\\
& D_\mu \tilde {\bar\beta} + \partial_{\theta} {\tilde {\bar {\cal F}}}_{\mu} = 0, \;\qquad
\partial_\theta \tilde {\Phi} + \partial_{\bar\theta} \tilde {\bar\beta } = 0, \; \qquad 
\partial_{\bar\theta} \tilde {\Phi} + \partial_\theta \tilde \beta  = 0, \;
\qquad \partial_{\bar\theta} \tilde \beta  = 0. &
\end{eqnarray}
It is evident that the above relations would provide us connection between the secondary fields of the expansions in equations (15)
and (23) and the basic as well as auxiliary fields of the ordinary 4D topologically massive non-Abelian gauge theory.

Let us take up the relations (i.e. $\partial_\theta \tilde {\Phi} + \partial_{\bar\theta} \tilde {\bar\beta } = 0,\;
\partial_{\bar\theta} \tilde {\Phi} + \partial_\theta \tilde \beta  = 0,\;
\partial_\theta \tilde {\bar\beta}  = 0,\; \partial_{\bar\theta} \tilde \beta  = 0$) and see their consequences. 
It turns out (cf. (27))  that we obtain the reduced form of   
$\tilde \beta, \tilde {\bar\beta} $ and $\tilde \Phi$ (with $b_1 = b_2 = b_3 =  f_2 = \bar f_3 = 0$ 
and $\bar f_1 + f_3 = 0, \; f_1 + \bar f_2 = 0$) as
\begin{eqnarray}
\tilde {\beta}^{(r)}(x,\theta, \bar \theta ) &=& \beta (x)+ \theta \; {\bar f}_2(x),
\;  \qquad \tilde {\bar\beta}^{(r)}(x,\theta, \bar \theta ) 
= \bar\beta (x)+ \bar\theta \; f_3 (x), \nonumber\\ 
\tilde \Phi^{(r)}(x,\theta, \bar \theta ) &=& \phi(x) + \theta \;(-\; f_3(x)) + \bar\theta \;(-\; \bar f_2(x)).
\end{eqnarray}
Making the choice $f_3 = -\; \bar f_1 = \rho,\quad f_1 = - \bar f_2 = \lambda$, we obtain 
\begin{eqnarray}
\tilde \beta ^{(r)}(x,\theta, \bar \theta ) &=& \beta(x) - \theta \;\lambda (x), \;
 \qquad \tilde {\bar\beta} ^{(r)}(x,\theta, \bar \theta ) 
= \bar\beta (x)+ \bar\theta\; \rho (x), \nonumber\\ 
\tilde \Phi^{(r)} (x,\theta, \bar \theta ) &=& \phi(x) - \theta\; \rho (x) + \bar\theta \;\lambda (x),
\end{eqnarray}
where $\rho (x) = \rho (x) \cdot T,\; \lambda (x) = \lambda (x) \cdot T$  and the superscript $(r)$ denotes the superfields that are 
obtained after the application of HC. In our further computations, we shall be exploiting (29) for the superfields
$\tilde \beta ,\tilde {\bar\beta}$ and $\tilde \Phi$ in their reduced form.

Now we take the relations $D_\mu \tilde \beta^{(r)} + \partial_{\bar\theta} \tilde {\cal F}_\mu = 0$ and 
$D_\mu \tilde {\bar\beta}^{(r)} + \partial_\theta {\tilde {\bar{\cal F}}}_\mu = 0$. These lead to: 
$B_\mu^{(1)} = -\;D_\mu \beta, \;S_\mu = i\; D_\mu \lambda,\; \bar B_\mu^{(2)} = -\; D_\mu \bar\beta, \; {\bar S}_\mu = i\; D_\mu \rho$.
The substitution of these makes $\tilde {\cal F}_\mu \rightarrow \tilde {\cal F}_\mu^{(r)}, 
\; {\tilde {\bar{\cal F}}}_\mu \rightarrow {\tilde {\bar{\cal F}}}_\mu^{(r)}$ as given bellow:
\begin{eqnarray}
\tilde {\cal F}_\mu^{(r)}(x,\theta, \bar \theta ) &=& C_\mu (x) + \theta \;{\bar B}_\mu^{(1)}(x) + \bar \theta \;[-\; D_\mu \beta (x)] 
+ \theta\;\bar\theta \;[-\; D_\mu \lambda (x) ], \nonumber\\
{\tilde {\bar{\cal F}}}_\mu^{(r)}(x,\theta, \bar \theta ) &=& {\bar C}_\mu (x) 
+ \theta \; [-\;D_\mu {\bar \beta} (x)] + \bar \theta \; B_\mu^{(2)} (x) 
+ \theta \;\bar\theta \;[-\; D_\mu \rho (x) ].
\end{eqnarray}
We focus now on the relationship $D_\mu \tilde \phi^{(r)} + \partial_\theta \tilde {\cal F}_{\mu}^{(r)}
 + \partial_{\bar\theta} {\tilde {\bar {\cal F}}}_{\mu}^{(r)} = 0$.
An explicit computation results in the following Curci-Ferrari (CF) type restriction 
\begin{eqnarray}
B_\mu^{(2)} + {\bar B}_\mu^{(1)} + D_\mu \phi = 0  
\qquad  \Longrightarrow   \qquad  B_\mu + {\bar B}_\mu + D_\mu \phi = 0,
\end{eqnarray}
where we have identified $B_\mu^{(2)} = B_\mu$ and ${\bar B}_\mu^{(1)} = {\bar B}_\mu$. 
Next, we take up the top two relations of equation (27) which can be expressed, as:
\begin{eqnarray}
& D_\mu \tilde {\cal F}_\nu^{(r)} - D_\nu \tilde {\cal F}_\mu^{(r)} + \partial_{\bar\theta} \tilde {\cal B}_{\mu\nu} 
+ i\; \Bigl [\tilde {\cal F}_{1}, F_{\mu\nu}\Bigr ] = 0, & \nonumber\\
& D_\mu {\tilde {\bar{\cal F}}}_\nu^{(r)} - D_\nu {\tilde {\bar{\cal F}}}_\mu^{(r)} + \partial_{\theta} \tilde {\cal B}_{\mu\nu} 
+ i\; \Bigl [{\tilde {\bar{\cal F}}}_{1}, F_{\mu\nu}\Bigr ] = 0.& 
\end{eqnarray}
Appropriate substitutions imply the following
\begin{eqnarray}
& R_{\mu\nu} = -\;( D_\mu C_\nu - D_\nu C_\mu ) + C_1 \times F_{\mu\nu}, \;\quad  
{\bar R}_{\mu\nu} = -\;( D_\mu {\bar C}_\nu - D_\nu {\bar C}_\mu ) + {\bar C}_1 \times F_{\mu\nu},& \nonumber\\
& S_{\mu\nu} = -i\;( D_\mu {\bar B}_\nu - D_\nu {\bar B}_\mu ) - {\bar R} \times F_{\mu\nu}  
\equiv i\;( D_\mu B_\nu - D_\nu B_\mu ) + S \times F_{\mu\nu}, & \nonumber\\
& R =i\;\beta, \qquad s_1 = i\;\lambda, \;
\qquad  {\bar S} = i\; \bar\beta, \; \qquad {\bar s}_1 = i\;\rho. &
\end{eqnarray}
It is very interesting to note that the equality of $S_{\mu\nu}$, in the above, produces another CF-type constraint 
that exists amongst the auxiliary fields and scalar field as
\begin{eqnarray}
& {\bar R} + S = i\;\phi \qquad \Longrightarrow  \qquad {\bar B_1} + B_1 = i\;\phi, & 
\end{eqnarray}
where we have identified ${\bar R} = {\bar B_1}$ and $S = B_1$. We have also used the CF-type  constraint 
(31) in the derivation of the above relationship from the equality of  $S_{\mu\nu}$ (cf. (33)).

Finally, we devote our attention to the comparison of the coefficients of the spacetime differentials $(dx^{\mu} \wedge dx^{\nu} \wedge dx^{\eta})$
from the l.h.s. and r.h.s. of the HC (i.e. $ \tilde {\cal H}^{(3)} = H^{(3)}$). The above requirement yields the following relationship
\begin{eqnarray}
& D_\mu \tilde {\cal B}_{\nu\eta}^{(r)} + D_\nu \tilde {\cal B}_{\eta\mu}^{(r)} + D_\eta  \tilde {\cal B}_{\mu\nu}^{(r)} 
+ i\;\Bigl [ \tilde {\cal K}_\mu, F_{\nu\eta} \Bigr ] + i\; \Bigl [ \tilde {\cal K}_\nu, F_{\eta\mu} \Bigr ] + 
i\; \Bigl [ \tilde {\cal K}_\eta, F_{\mu\nu} \Bigr ]& \nonumber\\ = & D_\mu B_{\nu\eta} + D_\nu  B_{\eta\mu} 
+ D_\eta  B_{\mu\nu} + i\;\Bigl [K_\mu, F_{\nu\eta} \Bigr ] + i\; 
\Bigl [ K_\nu, F_{\eta\mu} \Bigr ] + i\; \Bigl [ K_\eta, F_{\mu\nu} \Bigr ],&
\end{eqnarray}
where we have adopted the notation $D_\mu B_{\nu\eta} = \partial_\mu B_{\nu\eta} - A_\mu \times B_{\nu\eta}$, etc.
In the above, by exploiting relationships (33), it can be seen that
\begin{eqnarray}
\tilde {\cal B}_{\mu\nu}^{(r)}(x,\theta,\bar\theta) &=& B_{\mu\nu}(x) + \theta \; \Bigl [ -(D_\mu {\bar C}_\nu - D_\nu {\bar C}_\mu )
+ {\bar C}_1 \times F_{\mu\nu} \Bigr ]+ \bar\theta \; \Bigl [ -(D_\mu C_\nu - D_\nu C_\mu )\nonumber\\
&+& C_1 \times F_{\mu\nu} \Bigr ] + \theta \;\bar\theta \; \Bigl [ (D_\mu {\bar B}_\nu - D_\nu {\bar B}_\mu )
-i\;{\bar B_1} \times F_{\mu\nu} \Bigr ]. 
\end{eqnarray}
It is evident from (35) that the l.h.s. has terms that have coefficients of $\theta, \bar\theta$ and $\theta \bar\theta$. The r.h.s.,
however, has no such terms. Thus, these coefficients have to be set equal to zero. This restriction leads to the following 
relationships amongst the secondary fields of the expansion of $ \tilde {\cal K}^{(1)}(x,\theta, \bar \theta )$ (cf. (15) ) and the (anti-)ghost and
auxiliary fields, namely;
\begin{eqnarray}
P_\mu = D_\mu C_1 - C_\mu,  \quad {\bar P}_\mu = D_\mu {\bar C}_1 - {\bar C}_\mu, \quad
Q_\mu = - (D_\mu {\bar B_1} + i\; {\bar B}_\mu) \equiv (D_\mu B_1 + i\; B_\mu).
\end{eqnarray}
In the above comparison and derivation,  we have used  
\begin{eqnarray}
D_\mu F_{\nu\eta} + D_\nu F_{\eta\mu} + D_\eta F_{\mu\nu} = 0,
\end{eqnarray}
which is true for any arbitrary SU(N) gauge theory in any arbitrary dimension of spacetime.

The application of the horizontality type restrictions ($\tilde {\cal H}^{(3)} = H^{(3)}$) leads to 
the following super expansions for {\it all} the 
superfields of the theory:
\begin{eqnarray}
\tilde {\cal B}_{\mu\nu}^{(h)}(x,\theta,{\bar\theta}) &=& 
B_{\mu\nu} (x) + \theta \; \Bigl [ - \Bigl (D_\mu {\bar C}_\nu (x) - D_\nu {\bar C}_\mu (x) \Bigr)
+ {\bar C}_1 (x) \times F_{\mu\nu}(x) \Bigr ] \nonumber\\ &+&  
{\bar\theta} \;  \Bigl [ - \Bigl (D_\mu C_\nu (x) - D_\nu C_\mu (x) \Bigr )
+ C_1 (x) \times F_{\mu\nu} (x) \Bigr ] \nonumber\\ &+& 
\theta \;\bar\theta \; \Bigl [ \Bigl (D_\mu {\bar B}_\nu (x) - D_\nu {\bar B}_\mu (x) \Bigr )
-i\;{\bar B_1} (x) \times F_{\mu\nu} (x) \Bigr ] \nonumber\\ 
&\equiv & B_{\mu\nu}(x) + \theta\; \Bigl ( s_{ab}^{(2)} B_{\mu\nu}(x) \Bigr ) 
+ {\bar\theta} \; \Bigl (s_b^{(2)}B_{\mu\nu}(x) \Bigr ) 
+ \theta \;\bar\theta \;\Bigl( s_b^{(2)}s_{ab}^{(2)} B_{\mu\nu}(x)\Bigr), \nonumber\\
\tilde {\cal F}_\mu^{(h)}(x,\theta,{\bar\theta}) &=& C_\mu(x) + \theta \;{\bar B}_\mu (x) + {\bar\theta} \; \Bigl (-\;D_\mu \beta (x) \Bigr ) 
+ \theta \; \bar \theta \Bigl (- D_{\mu} \lambda (x) \Bigr )  \nonumber\\ 
& \equiv & C_\mu(x) + \theta \; \Bigl ( s_{ab}^{(2)}  C_\mu (x) \Bigr )
+ {\bar\theta} \; \Bigl ( s_{b}^{(2)}  C_\mu(x) \Bigr ) + \theta\;\bar\theta \; 
\Bigl ( s_b^{(2)}s_{ab}^{(2)}  C_\mu(x) \Bigr ), \nonumber\\
{\tilde {\bar {\cal F}}}_\mu^{(h)}(x,\theta,{\bar\theta}) &=& {\bar C}_\mu(x) 
+ \theta \; \Bigl (-\;D_\mu \bar\beta (x) \Bigr ) + \bar\theta \;B_\mu (x)
+ \theta\;\bar\theta \; \Bigl (-\; D_\mu \rho (x) \Bigr ) \nonumber\\ 
&\equiv & {\bar C}_\mu(x) + \theta \; \Bigl ( s_{ab}^{(2)} {\bar C}_\mu(x) \Bigr )
+ {\bar\theta} \; \Bigl ( s_{b}^{(2)} {\bar C}_\mu(x) \Bigr ) + \theta \;\bar\theta \; 
 \Bigl ( s_b^{(2)}s_{ab}^{(2)} {\bar C}_\mu(x) \Bigr ), \nonumber\\
\tilde \beta^{(h)}(x,\theta,{\bar\theta}) &=& \beta (x) + \theta \; \Bigl (-\; \lambda (x) \Bigr ) \nonumber\\
& \equiv&  \beta (x) + \theta \; \Bigl ( s_{ab}^{(2)} \beta (x)\Bigr ) + {\bar\theta} \; \Bigl ( s_{b}^{(2)} \beta (x)\Bigr ) 
+ \theta \;\bar\theta \; \Bigl ( s_b^{(2)}s_{ab}^{(2)} \beta (x)\Bigr ), \nonumber\\  
\tilde {\bar \beta}^{(h)}(x,\theta,\bar\theta ) &=& \bar \beta (x) + {\bar\theta} \; \rho (x)  \nonumber\\
& \equiv &  {\bar\beta} (x) + \theta \; \Bigl ( s_{ab}^{(2)} {\bar\beta} (x)\Bigr ) 
+ {\bar\theta}  \Bigl ( s_{b}^{(2)} {\bar\beta} (x)\Bigr ) 
+ \theta \;\bar\theta  \Bigl ( s_b^{(2)}s_{ab}^{(2)} \bar\beta (x)\Bigr ), \nonumber\\  
\tilde \Phi^{(h)}(x,\theta,\bar\theta ) &=& \phi(x) + \theta\; \Bigl (-\; \rho (x) \Bigr )
 + \bar\theta\; \lambda (x)  \nonumber\\   
& \equiv &  \phi(x) + \theta \; \Bigl ( s_{ab}^{(2)} \phi(x) \Bigr ) + \bar\theta\;  
\Bigl ( s_{b}^{(2)} \phi(x) \Bigr ) 
+ \theta \;\bar\theta \; \Bigl ( s_b^{(2)}s_{ab}^{(2)} \phi(x) \Bigr ), \nonumber\\
\tilde {\cal F}_1^{(h)}(x,\theta,\bar\theta ) &=& C_1(x) + \theta \; \Bigl (i\;{\bar B_1}(x) \Bigr ) 
+ \bar\theta \; \Bigr (-\; \beta (x) \Bigr ) 
+ \theta \;\bar\theta \; \Bigl (-\lambda (x) \Bigr ) \nonumber\\
& \equiv &  C_1(x) +\theta\; \Bigl ( s_{ab}^{(2)} C_1(x) \Bigr ) + \bar\theta\; \Bigl ( s_{b}^{(2)} C_1(x) \Bigr ) 
+ \theta \;\bar\theta\; \Bigl ( s_b^{(2)}s_{ab}^{(2)} C_1(x) \Bigr ),\nonumber\\
{\tilde {\bar{\cal F}}}_1^{(h)}(x,\theta,\bar\theta ) &=& {\bar C}_1(x) + \theta \; \Bigl (-\;{\bar\beta (x)} \Bigr ) 
+ \bar\theta \; \Bigl (i\;B_1 (x) \Bigr ) + \theta \;\bar\theta\; \Bigl (-\rho (x) \Bigr ) \nonumber\\
& \equiv &  {\bar C}_1(x) +\theta\; \Bigl ( s_{ab}^{(2)} {\bar C}_1(x) \Bigr ) + \bar\theta\; \Bigl ( s_{b}^{(2)} {\bar C}_1(x) \Bigr ) 
+ \theta \; \bar\theta \;\Bigl ( s_b^{(2)}s_{ab}^{(2)} {\bar C}_1(x) \Bigr ), \nonumber\\
\tilde {\cal K}_\mu^{(h)}(x,\theta,\bar\theta ) &=& K_\mu(x) + \theta \; 
\Bigl (D_\mu \bar C_1 (x) - \bar C_\mu (x) \Bigr ) 
+ \bar\theta \; \Bigl (D_\mu C_1 (x) - C_\mu (x) \Bigr ) \nonumber\\ 
&+&\theta\;\bar\theta \; \Bigl (i\;D_\mu B_1(x) - B_\mu (x) \Bigr ) \nonumber\\
& \equiv&  K_\mu(x) + \theta\; \Bigl ( s_{ab}^{(2)}  K_\mu(x)\Bigr ) 
+  \bar\theta\; \Bigl ( s_{b}^{(2)}  K_\mu(x)\Bigr ) 
+ \theta\;\bar\theta\; \Bigl ( s_b^{(2)}s_{ab}^{(2)}  K_\mu(x)\Bigr ),
\end{eqnarray}
where, in the above uniform expansions, we have taken into account some obvious transformations $ s_b^{(2)} \beta
 = 0,\; s_{ab}^{(2)} \bar\beta = 0,\; s_b^{(2)}s_{ab}^{(2)} \phi = 0$,
etc. Thus, the horizontality-type condition $(\tilde {\cal H}^{(3)} = H^{(3)})$ 
leads to the derivation of $\it all$ the (anti-)BRST symmetry transformations 
for all the basic fields $(B_{\mu\nu},\;C_\mu,\; {\bar C}_\mu,\; \beta, \; \bar\beta,\; \phi)$ 
as well as the compensating field $K_\mu(x)$ along with its associated ghost field $C_1(x)$ and anti-ghost field $ \bar C_1(x)$.\\

\begin{center}
{\section {(Anti-)BRST symmetry invariance: Curci-Ferrari type restrictions and coupled Lagrangian densities }}
\end{center}

It is evident, from the previous Sec. IV, that we have the following off-shell nilpotent 
and absolutely anticommuting (anti-)BRST symmetry transformations 
$s_{(a)b}^{(2)} $ corresponding to the``vector" gauge symmetry transformations $\delta_{v}$ (cf. (4)); namely;
\begin{eqnarray}
&&s_{b}^{(2)}B_{\mu\nu} = - \; (D_{\mu}C_{\nu} -D_{\nu}C_{\mu} ) + \; C_{1} \times F_{\mu\nu},\;\qquad
s_{b}^{(2)}C_{\mu} =- D_{\mu}\beta, \nonumber\\ &&s_{b}^{(2)}\bar C_{\mu} = B_{\mu},\;\qquad 
s_b^{(2)} {\bar B_1} = i\; \lambda,\;  \qquad s_b^{(2)} \bar C_1 = i\; B_{1},\;\qquad 
s_{b}^{(2)} \bar B_{\mu} = - D_{\mu} \lambda,
\nonumber\\ && s_{b}^{(2)}K_{\mu} =\; D_{\mu}C_{1} - C_{\mu},\;
\qquad s_{b}^{(2)} \phi = \lambda,\; \qquad s_b^{(2)} C_1 = -\beta,
\qquad s_{b}^{(2)} \bar \beta = \rho,\; \nonumber\\ && s_{b}^{(2)} [ A_{\mu},  F_{\mu\nu},
H_{\mu\nu\eta}, \beta,  B_1, \; \rho,  \lambda,  B_{\mu} ] = 0,
\end{eqnarray}
\begin{eqnarray}
&&s_{ab}^{(2)}B_{\mu\nu} = - \; (D_{\mu}\bar C_{\nu} -D_{\nu} \bar C_{\mu} ) + \; \bar C_{1} \times F_{\mu\nu},\; 
\qquad s_{ab}^{(2)}\bar C_{\mu} = - D_{\mu} \bar \beta, \nonumber\\
&& s_{ab}^{(2)}C_{\mu} = \bar B_{\mu},\; \qquad s_{ab}^{(2)}B_{\mu} = D_{\mu} {\rho},\;
\qquad s_{ab}^{(2)}C_1 = i\; \bar B_{1},\; \qquad s_{ab}^{(2)} \phi = -\rho, \nonumber\\
&& s_{ab}^{(2)} \bar C_1 = - \bar \beta, \; \qquad s_{ab}^{(2)}B_{1} = - i \; \rho, \;
\qquad s_{ab}^{(2)}K_{\mu} =\; D_{\mu} \bar C_{1} - \bar C_{\mu}, \nonumber\\ && s_{ab}^{(2)} \beta = - \lambda,
\qquad s_{ab}^{(2)} [A_{\mu}, F_{\mu\nu}, H_{\mu\nu\eta},
  \bar \beta,   \bar B_1,  \rho,  \lambda,  \bar B_{\mu} ] = 0.
\end{eqnarray}
It is interesting to check that the above (anti-)BRST transformations are absolutely anticommuting 
 on the constrained surface defined by the CF-type 
restrictions
\begin{eqnarray}
&& B + \bar B = -i\;(C \times \bar C),\; \qquad  B_1 + \bar B_1 = i\; \phi, \;
\qquad B_\mu + \bar B_\mu = - D_\mu \phi.  
\end{eqnarray}
For instance, it can be checked that $\{s_b^{(2)},\;s_{ab}^{(2)}\}B_{\mu\nu}=0$ and $\{s_b^{(2)},\;s_{ab}^{(2)}\}K_{\mu}=0$
if and only if the last two CF-type conditions, from the above, are satisfied.
Furthermore, it is pretty  straightforward to check that $s_{(a)b}^{(2)} [B_1+ \bar B_1 - i\;\phi]=0$,\quad $s_{(a)b}^{(2)} [B_\mu + \bar B_\mu +D_\mu\phi]=0$
under the above (anti-)BRST symmetry transformation (cf. (40) and (41)).

Exploiting the (anti-)BRST symmetry transformations from (40) and (41), we can derive the coupled Lagrangian 
densities for our present 4D theory as
\begin{eqnarray}
{\cal L}_{B_1} &=& - \frac {1}{4} F^{\mu\nu}\cdot F_{\mu\nu}+ \frac {1}{12}H^{\mu\nu\eta}\cdot H_{\mu\nu\eta}+\frac{m}{4}\varepsilon_{\mu\nu\eta\kappa}
B^{\mu\nu}\cdot F^{\eta\kappa} \nonumber\\
&+& s_b^{(2)} s_{ab}^{(2)} \Bigl [2\;\bar\beta \cdot \beta + \bar C_\mu \cdot C^\mu - \frac{1}{4} B^{\mu\nu} \cdot B_{\mu\nu}\Bigr ], \nonumber\\
{\cal L}_{\bar B_1} &=& - \frac {1}{4} F^{\mu\nu}\cdot F_{\mu\nu}+ \frac {1}{12}H^{\mu\nu\eta}\cdot H_{\mu\nu\eta}+\frac{m}{4}\varepsilon_{\mu\nu\eta\kappa}
B^{\mu\nu}\cdot F^{\eta\kappa} \nonumber\\ &-& s_{ab}^{(2)} s_b^{(2)} \Bigl [2\;\bar\beta \cdot \beta + \bar C_\mu \cdot C^\mu -
 \frac{1}{4} B^{\mu\nu} \cdot B_{\mu\nu}\Bigr ].
\end{eqnarray}
In general, one can choose, in the square brackets, terms like $(A_\mu \cdot A^\mu ), (\phi \cdot \phi )$. 
The first term, however, does not yield anything because 
$s_{(a)b}^{(2)} A_\mu = 0$  and the second term produces the same 
expression as one gets from $(\bar\beta \cdot \beta)$. Note that we have not
taken $(\bar C_1 \cdot C_1)$ in the above bracket because its mass dimension is zero whereas the mass dimension
of the other terms in the brackets (with ghost number zero) is two.
The explicit forms of the above Lagrangian densities, that respect the above 
(anti-)BRST symmetry transformations, are 
\begin{eqnarray}
{\cal L}_{B_1} &=& - \frac {1}{4} F^{\mu\nu}\cdot F_{\mu\nu}+ \frac {1}{12}H^{\mu\nu\eta}\cdot H_{\mu\nu\eta}+\frac{m}{4}\varepsilon_{\mu\nu\eta\kappa}
B^{\mu\nu}\cdot F^{\eta\kappa} + B^{\mu} \cdot B_\mu  \nonumber\\
&-& \frac{i}{2}\; B^{\mu\nu} \cdot (B_1 \times F_{\mu\nu}) - (D_\mu B^{\mu\nu} - D^{\nu}\phi) \cdot B_\nu 
+ D_\mu \bar \beta \cdot D^{\mu} \beta \nonumber\\ &+& \frac{1}{2} \;\Bigl [(D_\mu \bar C_\nu - D_\nu \bar C_\mu ) 
- \bar C_1 \times F_{\mu\nu} \Bigr ] \cdot \Bigl [(D^\mu C^\nu - D^\nu C^\mu ) - C_1 \times F^{\mu\nu} \Bigr ]\nonumber\\
&+& \rho \cdot (D_\mu C^{\mu} - \lambda ) +  (D_\mu {\bar C}^{\mu} - \rho )\cdot \lambda, \nonumber\\
{\cal L}_{\bar B_1} &=& - \frac {1}{4} F^{\mu\nu}\cdot F_{\mu\nu}+ \frac {1}{12}H^{\mu\nu\eta}\cdot H_{\mu\nu\eta}+\frac{m}{4}\varepsilon_{\mu\nu\eta\kappa}
B^{\mu\nu}\cdot F^{\eta\kappa} + {\bar B}^{\mu} \cdot {\bar B}_\mu  \nonumber\\
&+& \frac{i}{2}\; B^{\mu\nu} \cdot ({\bar B}_1 \times F_{\mu\nu}) + (D_\mu B^{\mu\nu} + D^{\nu}\phi) \cdot {\bar B}_\nu 
+ D_\mu \bar \beta \cdot D^{\mu} \beta \nonumber\\ &+& \frac{1}{2} \;\Bigl [(D_\mu \bar C_\nu - D_\nu \bar C_\mu ) 
- \bar C_1 \times F_{\mu\nu} \Bigr ] \cdot \Bigl [(D^\mu C^\nu - D^\nu  C^\mu ) - C_1 \times F^{\mu\nu} \Bigr ]\nonumber\\
&+& \rho \cdot (D_\mu C^\mu - \lambda ) +  (D_\mu {\bar C}^\mu - \rho )\cdot \lambda.
\end{eqnarray}
The transformations of the above Lagrangian densities under the (anti-)BRST transformations 
can be expressed in terms of the total spacetime derivatives as:
\begin{eqnarray}
s_{ab}^{(2)} {\cal L}_{\bar B_1} &=& - \; \partial_\mu \;\Bigl [ m\;\varepsilon^{\mu\nu\eta\kappa} {\bar C}_\nu \cdot (\partial_\eta A_\kappa -\;
\frac{1}{2}\; A_\eta \times A_\kappa ) + \rho \cdot {\bar B}^\mu + \lambda \cdot D^\mu {\bar \beta}\nonumber\\
&+& (D^\mu {\bar C}^\nu - D^\nu {\bar C}^\mu ) \cdot {\bar B}_\nu - ({\bar C}_1 \times F_{\mu\nu} ) \cdot {\bar B}_\nu \Bigr ], \nonumber\\
s_{b}^{(2)} {\cal L}_ {B_1} &=& + \; \partial_\mu \;
\Bigl [-\; m\;\varepsilon^{\mu\nu\eta\kappa} C_\nu \cdot (\partial_\eta A_\kappa -\;
\frac{1}{2}\; A_\eta \times A_\kappa ) + \rho\cdot D^\mu \beta + \lambda \cdot B^\mu \nonumber\\
 &+& (D^\mu C^\nu - D^\nu C^\mu ) \cdot B_\nu -  (C_1 \times F^{\mu\nu} ) \cdot  B_\nu \Bigr ].
\end{eqnarray}
This shows that the action integral, corresponding to the Lagrangian densities ${\cal L}_{\bar B_1}$ and ${\cal L}_{B_1}$, 
remains invariant under the (anti-)BRST symmetry transformations.

The above Lagrangian densities respect the (anti-)BRST symmetry transformations 
{\it together} because it can be checked that
$s_b^{(2)}{\cal L}_{\bar B_1}$ and $s_{ab}^{(2)}{\cal L}_{B_1}$ transform to the total spacetime derivatives plus terms that
turn out to be zero on the CF-type restrictions (42). 
Thus, we conclude that the Lagrangian densities ${\cal L}_{ B_1}$ and ${\cal L}_{\bar B_1}$
are equivalent on the constrained surface (in the Minkowskian spacetime manifold) described by the CF-type field equations.

We close this section with the comment that all the CF-type restrictions (cf. 42) are (anti-)BRST invariant as it can be checked, 
using (10), (40) and (41), that we have 
\begin{eqnarray}
s_{(a)b}^{(2)}\;[B_1 + {\bar B}_1 - i \; \phi ] = 0,\;\;
s_{(a)b}^{(2)} \; [B_\mu + {\bar B}_\mu + D_\mu \phi] = 0,\;\;
s_{(a)b}^{(1)}\;[B + \bar B + i\; (C \times \bar C)] = 0.
\end{eqnarray}
Thus, in some sense, the CF-type restrictions (42) are ``physical" conditions because they remain (anti-)BRST invariant under $s^{(2)}_{(a)b}$ and $s^{(1)}_{(a)b}$ and, hence, are gauge invariant.\\

\begin{center}
{\section{Invariance of the Curci-Ferrari type restrictions and Lagrangian densities: superfield formalism}}
\end{center}

First of all, we capture here the (anti-)BRST invariance of the CF-type restrictions. To this end in mind, we 
note that when we set the coefficient of $(dx^{\mu} \wedge d\theta \wedge d\bar\theta)$ equal to zero due to horizontality-type restriction $(\tilde{\cal H}^{(3)} = H^{(3)}),$ we obtain the following relationship:
\begin{eqnarray}
D_\mu \tilde \Phi^{(h)} + \partial _{\bar\theta} {\tilde {\bar {\cal F}}}_{\mu}^{(h)}
 + \partial_ {\theta} \tilde{\cal F}_{\mu}^{(h)} = 0,
\end{eqnarray}
where the superscript $(h)$ denotes the expression of the corresponding superfields obtained after the application of 
the HC (i.e. $\tilde{\cal H}^{(3)} = H^{(3)}$). The
above relationship leads to the emergence of CF-type restriction 
$B_\mu + \bar B_\mu + D_\mu \phi = 0$ from equation (47).

As we have noted earlier (cf. (11)), the (anti-)BRST symmetry  transformations $s_{(a)b}^{(2)}$ correspond to the translational generators along the 
Grassmannian directions (i.e. $s_{b}^{(2)}\rightarrow {\partial}/{\partial \bar\theta},\; s_{ab}^{(2)}\rightarrow {\partial}/{\partial \theta} )$,
the (anti-)BRST invariance of (31) can be proven by observing  
\begin{eqnarray}
\partial_{\theta}\; [D_\mu \tilde \Phi^{(h)} + \partial _{\bar\theta} {\tilde {\bar{\cal F}}}_{\mu}^{(h)} 
+ \partial_ {\theta} \tilde{\cal F}_\mu^{(h)}] = 0, \nonumber\\
\partial_{\bar\theta}\;[D_\mu \tilde \Phi^{(h)} + \partial _{\bar\theta} {\tilde {\bar{\cal F}}}_{\mu}^{(h)} 
+ \partial_ {\theta} \tilde {\cal F}_\mu^{(h)}] = 0.
\end{eqnarray}
Taking the help of expressions (39), it can be checked that 
\begin{eqnarray}
\partial_{\theta}\;D_\mu \tilde \Phi^{(h)} = -\;D_\mu \rho,\; \qquad  
\qquad \partial_{\theta}\partial_{\bar\theta}\;{\tilde {\bar{\cal F}}}^{(h)}_\mu 
= D_\mu \rho, \nonumber\\
\partial_{\bar\theta}\;D_\mu \tilde \Phi^{(h)} = D_\mu \lambda,\; \qquad
\qquad \partial_{\bar\theta}\partial_\theta\;\tilde {\cal F}^{(h)}_\mu 
=-\; D_\mu \lambda .
\end{eqnarray}
As a consequence, it is elementary to check that the equations (48) 
are trivially satisfied which, ultimately, imply the (anti-)BRST invariance:
$s_{(a)b}^{(2)}\;[B_\mu + \bar B_\mu + D_\mu \phi] = 0.$

To capture the CF-type restriction $B_1 + \bar B_1 = i\; \phi,$ within the framework of superfield formalism, is a bit tricky.
First of all, we note that when we set equal to zero the coefficients of $(dx^{\mu} \wedge dx^{\nu} \wedge d\theta)$ and  
$(dx^{\mu} \wedge dx^{\nu} \wedge d\bar\theta)$ in the equality $ \tilde {\cal H}^{(3)} = H^{(3)}$, we obtain 
\begin{eqnarray}
& D_\mu {\tilde {\bar{\cal F}}}_\nu^{(h)} - D_\nu {\tilde {\bar{\cal F}}}_\mu^{(h)}
 + \partial_\theta \tilde {\cal B}_{\mu\nu}^{(h)} + 
i\;[ {\tilde {\bar {\cal F}}}_1^{(h)}, F_{\mu\nu}] = 0,&
\end{eqnarray}
\begin{eqnarray}
& D_\mu \tilde {\cal F}_\nu^{(h)} - D_\nu \tilde {\cal F}_\mu^{(h)} + \partial_{\bar\theta}  \tilde {\cal B}_{\mu\nu}^{(h)} + 
i\;[\tilde{\cal F}_1^{(h)}, F_{\mu\nu}] = 0.&
\end{eqnarray}
It should be noted that 
1-form gauge field $A_\mu$ and the corresponding 2-form curvature tensor $F_{\mu\nu}$ do not transform under $s_{(a)b}^{(2)}$. 
As a consequence, we have taken $\tilde {\cal F}_{\mu\nu}^{(h)}(x, \theta,\bar\theta) = F_{\mu\nu}(x)$.
Applying $\partial_{\bar\theta}$ on (50) and $\partial_\theta$ on (51) and summing them up, leads to the following
\begin{eqnarray}
D_\mu ( \partial_\theta \tilde {\cal F}_\nu^{(h)} +\partial_{\bar\theta} {\tilde {\bar{\cal F}}}_{\nu}^{(h)}) - 
D_\nu ( \partial_\theta \tilde {\cal F}_\mu^{(h)} +\partial_{\bar\theta} {\tilde {\bar{\cal F}}}_{\mu}^{(h)}) + 
i\;[ \partial_\theta \tilde  {\cal F}_1^{(h)} + \partial_{\bar\theta} {\tilde  {\bar{\cal F}}}_1^{(h)},F_{\mu\nu}] = 0.
\end{eqnarray}
The substitution of the equation (47) yields the following relationship:
\begin{eqnarray}
\partial_\theta \tilde  {\cal F}_1^{(h)} + \partial_{\bar\theta} {\tilde {\bar{\cal F}}}_1^{(h)}= -\;\tilde \Phi^{(h)}.
\end{eqnarray}
The insertion  of the expressions for the superfields from (39), leads to the derivation of $B_1 + {\bar B}_1 = i\;\phi$.
It is now elementary to check that the
application of $\partial_\theta$ and $\partial_{\bar\theta}$ on (53) produces zero result.
As a consequence, it becomes clear that $s_{(a)b}^{(2)}\;(B_1 + {\bar B}_1 - i\;\phi)= 0$.

Now we discuss the (anti-)BRST invariance of the third CF-type restrictions
$ \Bigl (B + {\bar B} = -\;i\;(C \times {\bar C})\Bigr)$ within the framework of superfield formalism.
This restriction emerges from the following equality (that results in when we set equal to zero the coefficient of 
($ d\theta \wedge d\bar\theta$)  
\begin{eqnarray}
\partial_{\theta} \tilde {\cal F}^{(h)} + \partial_{\bar\theta} {\tilde {\bar {\cal F}}}^{(h)} -
i\;\lbrace \tilde {\cal F}^{(h)},{\tilde{\bar {\cal F}}}^{(h)} \rbrace = 0,
\end{eqnarray}
from the requirement of HC (cf. Sec. III). Exploiting the explicit form of the expansions (8), 
obtained after the application of the HC, namely;
\begin{eqnarray}
\tilde{\cal F}^{(h)} (x, \theta, \bar \theta ) &=& C(x) + \theta \; \Bigl( i\; {\bar B}(x)\Bigr ) 
+ \bar \theta \; \Bigl [\frac {1}{2}( C \times C ) (x) \Bigr ]  
+ \theta \;\bar \theta \; \Bigl [-i\;(\bar B \times C ) (x) \Bigr ], \nonumber\\ 
{\tilde {\bar {\cal F}}}^{(h)} (x, \theta, \bar\theta ) &=& {\bar C}(x) 
+ \theta \; \Bigl [\frac {1}{2} \; ( \bar C \times \bar C )(x)\Bigr ]  
+ \bar \theta \; (i\;B(x)) + \theta \;\bar \theta\;  \Bigl [-i\;( B \times \bar C ) (x) \Bigr ],
\end{eqnarray}
\noindent
it can be checked that the following relationships are true, namely;
\begin{eqnarray}
\partial_{\theta} \Bigl [\partial_{\theta} \tilde {\cal F}^{(h)} + \partial_{\bar\theta} {\tilde {\bar {\cal F}}}^{(h)} -
i\;\lbrace \tilde {\cal F}^{(h)},{\tilde{\bar {\cal F}}}^{(h)} \rbrace \Bigr ] = 0, \nonumber\\
\partial_{\bar\theta} \Bigl [\partial_{\theta} \tilde {\cal F}^{(h)} + \partial_{\bar\theta} {\tilde {\bar {\cal F}}}^{(h)} -
i\;\lbrace \tilde {\cal F}^{(h)},{\tilde{\bar {\cal F}}}^{(h)} \rbrace \Bigr ] = 0.
\end{eqnarray}
where the following inputs (obtained from (8)) play important roles:
\begin{eqnarray}
&&\partial_{\theta}\;\partial_{\bar \theta} {\tilde {\bar {\cal F}}}^{(h)} = + i\;(B \times {\bar C}),\;\qquad \quad  
\partial_{\theta} \Bigl [-\;i\;\lbrace \tilde {\cal F}^{(h)},{\tilde{\bar {\cal F}}}^{(h)} \rbrace \Bigr ] 
= -i\; (B \times {\bar C}),\nonumber\\
&&\partial_{\bar\theta}\;\partial_{\theta} \tilde {\cal F}^{(h)} = -i\;({\bar B} \times C),\;\qquad \quad
\partial_{\bar \theta} \Bigl [-\;i\;\lbrace \tilde {\cal F}^{(h)},{\tilde{\bar {\cal F}}}^{(h)} \rbrace \Bigr ] 
=+i\; ({\bar B} \times C).
\end{eqnarray}
It can be verified explicitly that the relations (56) are true. With our observations of $s_{b}^{(1)}\rightarrow  {\partial}/{\partial {\bar\theta }},\; s_{ab}^{(1)}\rightarrow {\partial}/{\partial \theta} $ (cf.  Sec. III), 
it is evident that the CF-condition $(B + {\bar B} = -i\;(C \times {\bar C}))$
is an (anti-)BRST invariant relationship
(i.e. $\;s_{(a)b}^{(1)} [ B + {\bar B } + i\;(C \times {\bar C})] = 0$).

We discuss now the (anti-)BRST invariance of the coupled Lagrangian densities of the theory within the framework of
the superfield formulation. It is essential to recall that the kinetic term of the 1-form gauge field can be expressed in terms of superfield
as : $\Bigl (-\;\frac {1}{4} F^{\mu \nu}(x) \cdot  F_{\mu \nu}(x) 
=  -\;\frac {1}{4} \tilde {\cal F}^{\mu \nu (h)}(x,\theta,\bar\theta ) \cdot  
\tilde {\cal F}_{\mu \nu}^{(h)}(x,\theta,\bar \theta ) \Bigr )$
because of the horizontality condition (cf. Sec. III). In exactly similar fashion, the kinetic term for the 2-form gauge field can be expressed as:
$\Bigl (\frac {1}{12} H^{\mu\nu\eta}(x) \cdot  H_{\mu \nu\eta}(x) 
=  \frac {1}{12} \tilde{\cal H}^{\mu \nu \eta( h)}(x,\theta,\bar\theta ) \cdot  
\tilde{\cal H}_{\mu \nu \eta}^{(h)}(x,\theta,\bar \theta ) \Bigr )$
because of our discussion in Sec. IV. It may be mentioned here that the kinetic terms are actually independent of the Grassmannian variables
because of the application of HCs (as is evident from the above equalities). Thus, they are automatically (anti-)BRST invariant.

The topological mass term can be expressed in terms of the superfields, as 
\begin{eqnarray}
T(x) = \frac {m}{4}\;\varepsilon^{\mu\nu\eta\kappa}
\;B_{\mu\nu}(x)\cdot F_{\eta\kappa}(x)\longrightarrow \tilde {\cal T}(x,\theta,\bar\theta) = \frac{m}{4}\;\varepsilon^{\mu\nu\eta\kappa}
\; \tilde {\cal B}_{\mu\nu}^{(h)}(x,\theta, \bar \theta )\cdot F_{\eta\kappa}(x).
\end{eqnarray}
After the substitution of the expansions from (39) for $\tilde {\cal B}_{\mu\nu}^{(h)}(x,\theta,\bar\theta)$,
it can be checked that the operation of $\partial_{\theta}$ and/or $\partial_{\bar\theta}$ 
on $\tilde {\cal T}(x,\theta,\bar\theta)$
always produces a total derivative term. Mathematically, this can be stated in the following  fashion:
\begin{eqnarray}
\lim_{\theta \to 0} \; \frac{\partial}{\partial \bar\theta}\; \tilde {\cal T} (x, \theta, \bar\theta) 
= - m \;\partial _\mu \;\Bigl [\varepsilon^{\mu\nu\eta\kappa} C_\nu \cdot (\partial_\eta A_\kappa -\frac{1}{2}\;
A_\eta \times A_\kappa ) \Bigr ], \nonumber\\
\lim_{\bar\theta \to 0} \; \frac{\partial}{\partial \theta}\; \tilde {\cal T}(x, \theta, \bar\theta ) 
= - m \;\partial _\mu \;\Bigl [\varepsilon^{\mu\nu\eta\kappa} {\bar C}_\nu \cdot (\partial_\eta A_\kappa -\frac{1}{2}\;
A_\eta \times A_\kappa ) \Bigr ], \nonumber\\ \frac{\partial}{\partial \bar\theta}
\;\frac{\partial}{\partial\theta} \tilde {\cal T}(x, \theta, \bar\theta) 
= + m \;\partial_\mu \; \Bigl [\varepsilon^{\mu\nu\eta\kappa} {\bar B}_\nu \cdot (\partial_\eta A_\kappa -\frac{1}{2}\;
A_\eta \times A_\kappa ) \Bigr ], 
\end{eqnarray}
where we have used the following results
\begin{eqnarray}
\varepsilon_{\mu\nu\eta\kappa} (C_1\times F^{\mu\nu})\cdot F^{\eta\kappa} = 0,
\quad \varepsilon_{\mu\nu\eta\kappa} ({\bar C}_1\times F^{\mu\nu})\cdot F^{\eta\kappa} = 0,\quad
\varepsilon_{\mu\nu\eta\kappa} ({\bar B}_1\times F^{\mu\nu})\cdot F^{\eta\kappa} = 0, 
\end{eqnarray}
due to specific property of $\varepsilon_{\mu\nu\eta\kappa}$ and totally antisymmetric nature of $f^{abc}$.
We conclude that the kinetic terms of 1-form and 2-form gauge fields and topological term of the coupled Lagrangian densities remain
(anti-)BRST invariant (modulo a total spacetime derivative).

The total Lagrangian densities (43) can now be generalized onto the (4, 2)-dimensional supermanifold 
after the application of the HC $(\tilde {\cal H}^{(3)} = H^{(3)})$, as 
\begin{eqnarray}
\tilde {\cal L}_{B_1} &=& - \; \frac {1}{4}\;\tilde {\cal F}^{\mu\nu (h)}\cdot \tilde{\cal F}_{\mu\nu}^{(h)}
+  \; \frac {1}{12}\;\tilde {\cal H}^{\mu\nu\eta (h)}\cdot \tilde{\cal H}_{\mu \nu\eta}^{(h)}
+ \frac {m}{4}\;\varepsilon_{\mu\nu\eta\kappa}\; (\tilde {\cal B}^{\mu\nu (h)}\cdot  F^{\eta\kappa})\nonumber\\ 
&+& \frac{\partial}{\partial \bar\theta}\;\frac{\partial}{\partial\theta} \;\Bigl [2\;{\tilde {\bar {\beta}}}^{(h)} \cdot \tilde \beta^{(h)} 
+ {\tilde {\bar {\cal F}}}_\mu^{(h)} \cdot \tilde { {\cal F}}^{\mu (h)} -\frac{1}{4} \tilde {\cal B}^{\mu\nu (h)} \cdot \tilde {\cal B}_{\mu\nu}^{(h)}\Bigr ],
\end{eqnarray}
\begin{eqnarray}
\tilde {\cal L}_{\bar B 1} &=& - \; \frac {1}{4}\;\tilde {\cal F}^{\mu\nu (h)} \cdot \tilde {\cal F}_{\mu\nu}^{(h)}
+  \; \frac {1}{12}\;\tilde {\cal H}^{\mu\nu\eta (h)}\cdot \tilde{\cal H}_{\mu \nu\eta}^{(h)}
+ \frac {m}{4}\;\varepsilon_{\mu\nu\eta\kappa}\; (\tilde {\cal B}^{\mu\nu (h)}\cdot F^{\eta\kappa})\nonumber\\ 
&-& \frac{\partial}{\partial\theta} \frac {\partial}{\partial \bar\theta} \;\Bigl [2\;\tilde {\bar\beta}^{(h)} \cdot \tilde \beta^{(h)} 
+ {\tilde {\bar {\cal F}}}_\mu^{(h)} \cdot \tilde { {\cal F}}^{\mu (h)} - \frac {1}{4} \tilde {\cal B}^{\mu\nu (h)} \cdot \tilde {\cal B}_{\mu\nu}^{(h)}\Bigr ].
\end{eqnarray}
From our earlier discussions, it is clear that the operation of the generators ${\partial}/{\partial\theta}$ 
and ${\partial}/{\partial \bar\theta}$
on the above expressions produces the zero result because of the nilpotency property $( \partial_\theta^{2} = 0, \partial_{\bar\theta}^{2} = 0 )$ and the
anticommutativity property $(\partial_\theta \;\partial_{\bar \theta} + \partial_{\bar\theta}\; \partial_\theta = 0 )$ 
of the translational generators  $\partial_\theta$ and $\partial_{\bar \theta}$. This observation captures the (anti-)BRST invariance
 of the coupled Lagrangian densities (cf. (43)-(45)) in the language of the superfield formulation because we observe that: $\partial_\theta \tilde {\cal L}_{B_1} = 0, \; \partial_{\bar \theta} \tilde {\cal L}_{B_1} = 0,\;
\partial_\theta\tilde {\cal L}_{\bar B_1} = 0, \; \partial_{\bar \theta} \tilde {\cal L}_{\bar B_1} = 0$.

We close this section with the remarks that the CF-type restrictions, 
that emerge  in our superfield formulation, are always (anti-)BRST invariant and 
they turn out to be responsible for the validity of anticommutativity property of the 
(anti-)BRST symmetry transformations as well as the subtled existence of the equivalent,
coupled and (anti-)BRST invariant Lagrangian densities
for the theory. The existence of the CF-type restriction and their connections
with the concept of gerbes are the specific features of our superfield approach
 to BRST formalism. Geometrically, the (anti-)BRST invariances are the CF-type restrictions as well as 
the coupled Lagrangian densities are the specific collection of superfields (obtained after the HC 
and the super derivatives on them) such that their translation along the Grassmannian directions 
$\theta$ and/or $\bar\theta$ produces the zero result.\\

\begin{center}
{\section{(Anti-)BRST symmetries corresponding to ``scalar" and ``vector" gauge symmetries: a bird's eye view}}
\end{center}

The off-shell nilpotent set of our (anti-)BRST symmetry transformations $(s_{(a)b}^{(1)}, s_{(a)b}^{(2)})$ 
[corresponding to the ``scalar" gauge symmetry transformation
$(s_{(a)b}^{(1)})$ and ``vector" gauge symmetry transformation $(s_{(a)b}^{(2)})$] obey the following algebra 
in their operator form:
\begin{eqnarray}
& (s_b^{(1)})^2 = (s_{ab}^{(1)})^2 = 0,\; \qquad \qquad (s_b^{(2)})^2 = (s_{ab}^{(2)})^2 = 0, & \nonumber\\
& s_b^{(1)}s_{ab}^{(1)} + s_{ab}^{(1)}s_b^{(1)} = 0 , \qquad \qquad
s_b^{(2)}s_{ab}^{(2)} + s_{ab}^{(2)}s_b^{(2)} = 0, &\nonumber\\
& s_b^{(1)}s_{ab}^{(2)} + s_{ab}^{(1)}s_b^{(2)}\ne 0 ,\qquad  \qquad s_b^{(1)}s_b^{(2)} + s_b^{(2)}s_b^{(1)} \ne 0 ,& \nonumber\\
& s_{ab}^{(1)}s_{ab}^{(2)} + s_{ab}^{(2)}s_{ab}^{(1)} \ne 0,\; \qquad  \qquad s_{ab}^{(1)}s_b^{(2)} + s_b^{(2)}s_{ab}^{(1)} \ne 0.&
\end{eqnarray}
As a consequence, even though, separately and independently, the 
(anti-)BRST transformations $(s_{(a)b}^{(1)}$ and $ s_{(a)b}^{(2)})$
are off-shell nilpotent as well as absolutely anticommuting, they do {\it not} anticommute with one-another 
(i.e. $s_{(a)b}^{(1)}\;s_{(a)b}^{(2)} + s_{(a)b}^{(2)}\;s_{(a)b}^{(1)} \neq 0).$

We can merge together the above off-shell nilpotent (anti-)BRST symmetries to generate 
a new set of off-shell nilpotent (anti-)BRST symmetry
transformations. The ensuing ``merged" off-shell nilpotent BRST symmetry transformations are as follows; 
\begin{eqnarray}
&& s_b B_{\mu\nu} = -\;(B_{\mu\nu} \times C) - (D_\mu C_\nu - D_\nu C_\mu ) + C_1 \times F_{\mu\nu},\;
\quad s_b {\bar C} = i\;B, \nonumber\\
&& s_b K_\mu = -\;(K_\mu \times C) + D_\mu C_1 - C_\mu, \;
\qquad s_b C_\mu = -\;D_\mu \beta + (C_\mu \times C), \nonumber\\
&& s_b {\bar C}_\mu = B_\mu,\quad s_b C = \frac{1}{2} (C \times C),\;
\quad s_b A_\mu = D_\mu C,\; \qquad s_b \beta = -\;(\beta \times C),\nonumber\\ 
&& s_b H_{\mu\nu\eta} = -\;(H_{\mu\nu\eta} \times C),\;
\quad s_b F_{\mu\nu} = -\;(F_{\mu\nu} \times C),\; \qquad  s_b \lambda = (\lambda \times C ),  \nonumber\\
&& s_b \bar C_1 = i\;B_1, \qquad s_b \phi = \lambda - (\phi \times C),\;
\qquad s_b \bar B_\mu = - D_\mu \lambda -(\bar B_\mu \times C ) , \nonumber\\
&& s_b {\bar B} = -\;({\bar B} \times C) ,\; \qquad s_b \bar\beta = \rho ,
\qquad s_b C_1 = - \beta + (C_1 \times C ), \nonumber\\
&& s_b {\bar B}_1 = i \;\lambda - ({\bar B}_1 \times C ),
\qquad s_b \Bigl[\rho, B, B_1,  B_\mu \Bigr ] = 0.
\end{eqnarray}
It should be carefully noted that the top two transformations, 
in the above, have been obtained due to the straightforward
sum $ s_b = s_b^{(1)} + s_b^{(2)}$ as the fields $B_{\mu\nu}$
and $K_\mu$ transform under both $s_b^{(1)}$ as well as $s_b^{(2)}$. The rest of the transformations in (64) have been obtained by the
requirements of the off-shell nilpotency of $s_b B_{\mu\nu}$ and $s_b K_\mu$. It can be explicitly checked that the above BRST transformations are off-shell
nilpotent (i.e. $s_b^{2} \psi = 0$ for $ \psi = B_{\mu\nu}, K_\mu, C_\mu, {\bar C}_\mu, B_\mu, {\bar B}_\mu, C_1, {\bar C}_1, C, {\bar C}, \beta, \bar\beta,
\phi, F_{\mu\nu}, H_{\mu\nu\eta}, B_1, {\bar B}_1, B, {\bar B}, \lambda, \rho $).

In exactly  similar fashion, one can merge the anti-BRST symmetry transformations $s_{ab}^{(1)}$ and  $s_{ab}^{(2)}$
to generate a new set of
off-shell nilpotent $(s_{ab}^{2} = 0)$ anti-BRST transformations $s_{ab}$. These anti-BRST symmetry transformations are as follows:
\begin{eqnarray}
&& s_{ab} B_{\mu\nu} = -\;(B_{\mu\nu} \times {\bar C}) - (D_\mu {\bar C}_\nu - D_\nu {\bar C}_\mu )
+ {\bar C}_1 \times F_{\mu\nu},\; \qquad s_{ab} C = i\;{\bar B}, \nonumber\\
&& s_{ab} F_{\mu\nu} = -\;(F_{\mu\nu} \times {\bar C}),\; \qquad
s_{ab} C_\mu = {\bar B}_\mu, \; \qquad
s_{ab} {\bar C}_\mu = -\;D_\mu \bar\beta + ({\bar C}_\mu \times {\bar C}), \nonumber\\ 
&& s_{ab} \bar\beta = -\;(\bar\beta \times {\bar C}),\;
\qquad s_{ab} \bar C = \frac{1}{2} (\bar C \times \bar C),  \; 
\quad s_{ab} B_1 = - i \;\rho - (B_1 \times {\bar C} ),  \nonumber\\ 
&& s_{ab} A_\mu = D_\mu \bar C , \;\quad 
s_{ab} \phi = -\;\rho - (\phi \times {\bar C}),\; \qquad s_{ab} B_\mu = D_\mu \rho -(B_\mu \times {\bar C} ),
\nonumber\\ 
&&  s_{ab} C_1 = i\;{\bar B}_1, \; \quad s_{ab} \beta = -\;\lambda, \; 
\qquad s_{ab} \rho = (\rho \times {\bar C}),\; \qquad s_{ab} B = -\;(B \times {\bar C}), \nonumber\\
&& s_{ab} K_\mu = -\;(K_\mu \times {\bar C}) + D_\mu {\bar C}_1 - {\bar C}_\mu, \;
 \qquad s_{ab} \bar C_1 = -\; \bar\beta 
+ ({\bar C}_1 \times {\bar C}), \nonumber\\
&& s_{ab} H_{\mu\nu\eta} = -\;(H_{\mu\nu\eta} \times {\bar C}),\; \qquad
s_{ab} \Bigl[\; \lambda, \; {\bar B},\; {\bar B}_1,\; {\bar B}_\mu \Bigr ] = 0.
\end{eqnarray}
It can be explicitly checked that, for the generic field $\psi$, we have $s_{ab}^{2} \psi = 0$. This 
establishes that the above (anti-)BRST transformations are indeed off-shell nilpotent transformations.

The above ``merged" (anti-)BRST transformations $s_{(a)b}$ are, however, not absolutely anticommuting 
(i.e. $s_b \; s_{ab} + s_{ab} \; s_b \ne 0 $). Thus,
these transformations are {\it not} linearly independent transformations. We 
can not obtain the coupled and {\it equivalent} Lagrangian densities that respect
the above (anti-)BRST symmetries $s_{(a)b}$
{\it together}. One can derive, however, the following BRST invariant Lagrangian  density:
\begin{eqnarray}
{\cal L}_B &=& - \frac{1}{4}\; F^{\mu\nu} \cdot F_{\mu\nu} + \frac{1}{12} \;H^{\mu\nu\eta} \cdot H_{\mu\nu\eta}
+ \frac{m}{4} \;\varepsilon_{\mu\nu\eta\kappa} B^{\mu\nu} \cdot F^{\eta\kappa} 
+ s_b s_{ab} \;\Bigl ( \frac{i}{2}\; A_\mu \cdot A^\mu \nonumber\\ &+& C \cdot \bar C + \frac{1}{2}\; \phi \cdot \phi
+ 2\; \bar \beta \cdot \beta + \bar C_\mu \cdot C^\mu + B_1 \cdot {\bar B}_1
- \frac{1}{4}\; B^{\mu\nu} \cdot B_{\mu\nu} \Bigr ).
\end{eqnarray}
It is very important to note that the above terms in the parenthesis possess mass dimension equal to two and they
carry the ghost number equal to zero. Furthermore, these terms provide the full gauge-fixing and Faddeev-Popov
ghost terms for the theory. It is evident that the above Lagrangian density would respect the off-shell
nilpotent BRST symmetry transformations (64) because of the off-shell nilpotency (i.e. $s_b^{2}$) of  $s_b$ 
and invariance of the kinetic and topological mass terms of the Lagrangian density ${\cal L}_B$ under $s_b$.
We have not taken the ghost number zero
combination $(\bar C_1 \cdot C_1)$ in the above parenthesis because it has the mass dimension equal to
zero in the natural units (where $\hbar = c = 1$).

To check the BRST invariance explicitly, the above BRST invariant Lagrangian density (66)
can be written, in its full blaze of glory, as
\begin{eqnarray}
{\cal L}_{B} = {\cal L}_{(0)} + {\cal L}_{(1)} + {\cal L}_{(2)},
\end{eqnarray}
where $ {\cal L}_{(0)}$ is the starting Lagrangian density (1) and  $ {\cal L}_{(1)}$ is the Lagrangian density that 
does not incorporate any type of (anti-)ghost fields, as given below:
\begin{eqnarray}
{\cal L}_{(1)} &=& B \cdot (\partial_\mu A^\mu) + \frac{1}{2} (B \cdot B + \bar B \cdot \bar B) + B^{\mu\nu} \cdot 
(D_\mu B_\nu - \frac{i}{2} B_1 \times F_{\mu\nu}) \nonumber\\
&-&\; i \; \bar B_1 \cdot (B_1 \times B ) - B_\mu \cdot \bar B^\mu.
\end{eqnarray}
The Lagrangian density that incorporates the (anti-)ghost fields is given by
\begin{eqnarray}
{\cal L}_{(2)} &=& -\; i\; \partial_\mu \bar C \cdot D^\mu C - B^{\mu\nu} \cdot \Bigl [ D_\mu C \times \bar C_\nu 
+ \frac{1}{2} \bar C_1 \times (F_{\mu\nu} \times C) \Bigr ] \nonumber\\  &-& \frac {1}{2}\;\Bigl [(B^{\mu\nu} \times C) 
+(D^\mu C^\nu -D^\nu C^\mu) - C_1 \times F_{\mu\nu}\Bigr ] \cdot \Bigl [ (D_\mu \bar C_\nu \nonumber\\
&-& D_\nu \bar C_\mu ) - \bar C_1 \times F_{\mu\nu} \Bigr ] + \rho \cdot \Big [ D_\mu C^\mu - \lambda - (\phi \times C) 
- i\;(\bar B_1 \times C) \nonumber\\ &+& 2\;(\beta \times \bar C) \Bigr ] + \Bigl [ D_\mu \bar C^\mu - \rho 
+ i\; (B_1 \times \bar C) - 2\;(\bar \beta \times C) \Bigr ] \cdot \lambda \nonumber\\
&-& (B_1 \times \bar C) \cdot (\bar B_1 \times C) + \Bigl [D_\mu \bar\beta - \bar C_\mu \times \bar C \Bigr ] \cdot 
\Bigl [D^\mu \beta - C^\mu \times C \Bigr ]\nonumber\\
&+& \Bigl [(D_\mu C \times \bar \beta ) + (B_\mu \times \bar C) - i\; (\bar C_\mu \times B ) \Bigr ]\cdot C^\mu 
- \bar C^\mu \cdot (\bar B_\mu \times C)\nonumber\\
&-& 2 \; (\bar \beta \times \bar C ) \cdot (\beta \times C) - 2\;i\; (\bar \beta \times B) \cdot \beta.
\end{eqnarray}
It can be checked (with the help of a bit of involved algebra) that the above Lagrangian density $ {\cal L}_B$
transforms under the BRST transformations (64) as
\begin{eqnarray}
 s_b \; {\cal L}_B &=& \partial_\mu \Bigl [ -\; m \; \varepsilon ^{\mu\nu\eta\kappa} C_\nu \cdot (\partial_\eta A_\kappa
- \frac {1}{2}\; A_\eta \times A_\kappa ) + B \cdot D^\mu C + B^\mu \cdot \lambda \nonumber\\ &+& \rho \cdot D^\mu \beta 
- \rho \cdot (C^\mu \times C) - \bar C^\mu \cdot (\lambda \times C) \Bigr ].
\end{eqnarray}
This shows that the action, corresponding to the above Lagrangian density ${\cal L}_B$, remains invariant under the BRST symmetry transformations (64).
In an exactly similar fashion, one can write the following Lagrangian density ${\cal L}_{\bar B}$
\begin{eqnarray}
 {\cal L}_{\bar B} &=& - \frac{1}{4}\; F^{\mu\nu} \cdot F_{\mu\nu} + \frac{1}{12} \;H^{\mu\nu\eta} \cdot H_{\mu\nu\eta}
+ \frac{m}{4} \;\varepsilon_{\mu\nu\eta\kappa} B^{\mu\nu} \cdot F^{\eta\kappa} 
- s_{ab} s_b \;\Bigl ( \frac{i}{2}\; A_\mu \cdot A^\mu \nonumber\\ &+& C \cdot {\bar C} + \frac{1}{2}\; \phi \cdot \phi 
+ 2 \; \bar\beta \cdot \beta + \bar C_\mu \cdot C^\mu 
+ B_1 \cdot {\bar B}_1 - \frac{1}{4}\; B^{\mu\nu} \cdot B_{\mu\nu} \Bigr ),
\end{eqnarray}
that would respect the nilpotent anti-BRST symmetry transformations $s_{ab}$ (cf. (65)) because these symmetries
are off-shell nilpotent (i.e. $s_{ab}^2 = 0$) and the kinetic and topological terms of the theory are gauge and anti-BRST
invariant by construction.

To check the anti-BRST invariance explicitly, it can be seen that the above Lagrangian density (71)
can be clearly written, in three parts, as given below
\begin{eqnarray}
{\cal L}_{\bar B} = {\cal L}_{(0)} + {\cal L}_{(b)} + {\cal L}_{(c)},
\end{eqnarray}
where ${\cal L}_{(0)}$ is the starting Lagrangian density (1) and ${\cal L}_{(b)}$ is the part that does not incorporate the 
(anti-)ghost fields. The latter can be written as
\begin{eqnarray}
{\cal L}_{(b)} &=& -\;\bar B \cdot (\partial_\mu A^\mu) + \frac{1}{2} (B \cdot B + \bar B \cdot \bar B) - B^{\mu\nu} \cdot 
\Bigl [ D_\mu \bar B_\nu - \frac{i}{2} (\bar B_1 \times F_{\mu\nu}) \Bigr ] \nonumber\\
&+&\; i \;  B_1 \cdot (\bar B_1 \times \bar B ) - B_\mu \cdot \bar B^\mu.
\end{eqnarray}
The Lagrangian density ${\cal L}_{(c)}$ in (72) contains all types of (anti-)ghost fields as given below
\begin{eqnarray}
{\cal L}_{(c)} &=&i\; \partial_\mu C \cdot D^\mu \bar C + B^{\mu\nu} \cdot \Bigl [ D_\mu \bar C \times C_\nu 
+ \frac{1}{2} C_1 \times (F_{\mu\nu} \times \bar C )\Bigr ] \nonumber\\ &+& \frac {1}{2}\;\Bigl [(B^{\mu\nu} \times \bar C) 
+(D^\mu \bar C^\nu -D^\nu \bar C^\mu) - \bar C_1 \times F_{\mu\nu}\Bigr ] \cdot \Bigl [ (D_\mu C_\nu \nonumber\\
&-& D_\nu C_\mu ) - C_1 \times F_{\mu\nu} \Bigr ] + \rho \cdot \Bigl [ D_\mu C^\mu - \lambda 
+ 2\; (\beta \times \bar C )   
- i\;(\bar B_1 \times C) \Bigr ] \nonumber\\ &+& \Bigl [ D_\mu \bar C^\mu - \rho 
+ i\; (B_1 \times \bar C) - 2\;(\bar \beta \times C) + (\phi \times \bar C) \Bigr ] \cdot \lambda \nonumber\\
&-& (B_1 \times \bar C) \cdot (\bar B_1 \times C) + (D_\mu \bar\beta - \bar C_\mu \times \bar C) \cdot (D^\mu \beta - C^\mu \times C)\nonumber\\
&-& \bar C_\mu \cdot \Bigl [ (D^\mu \bar C \times \beta ) + (\bar B^\mu \times C) - i\; ( C^\mu \times \bar B ) \Bigr ]
+ ( B_\mu \times \bar C)\cdot C^\mu  \nonumber\\
&-& 2 \; (\bar \beta \times \bar C ) \cdot (\beta \times C) + 2\;i\; \bar \beta \cdot( \beta \times \bar B).
\end{eqnarray}
With some involved algebra, it can be checked that, under the anti-BRST transformations (65), we have the 
following transformation for the Lagrangian density, namely;
\begin{eqnarray}
s_{ab} \; {\cal L}_{\bar B} &=& -\; \partial_\mu \Bigl [ \; m \; \varepsilon ^{\mu\nu\eta\kappa} \bar C_\nu \cdot (\partial_\eta A_\kappa
 - \frac {1}{2}\; A_\eta \times A_\kappa ) + \bar B \cdot D^\mu \bar C + \rho \cdot \bar B^\mu \nonumber\\ &+& \lambda \cdot D^\mu \bar \beta 
- \rho \cdot (C^\mu \times \bar C) - \bar C^\mu \cdot (\lambda \times \bar C) \Bigr ],
\end{eqnarray}
which shows that the action, corresponding to the Lagrangian density ${\cal L}_{\bar B}$, remains invariant under (65) because ${\cal L}_{\bar B}$
transforms to a total spacetime derivative.

We re-emphasize that the above Lagrangian densities ${\cal L}_B $ and ${\cal L}_{\bar B}$ are {\it not} equivalent 
because they do not respect the off-shell nilpotent (anti-)BRST symmetry transformations {\it together} as a consequence of the fact that 
(i.e. $s_b s_{ab} + s_{ab} s_b \neq 0$). We, at present, do not have any clue as to how to obtain the
$s_b$ and $s_{ab}$ that respect the absolute anticommutativity property. The lack of absolute anticommutativity 
(i.e. $s_b s_{ab} + s_{ab} s_b \neq 0$) owes its origin to the fact that these 
transformations have not been obtained by exploiting our superfield formulation. Rather, 
these transformations have been derived solely by the requirements of the off-shell nilpotency of the 
above (anti-)BRST symmetry transformations. \\

\begin{center}
{\section{Conclusions}}
\end{center}

We have obtained, in our present investigation, the off-shell nilpotent and absolutely anticommuting 
(anti-)BRST symmetry transformations corresponding to the ``scalar" and ``vector" local
gauge symmetry transformations of the topologically massive 4D non-Abelian 2-form gauge 
theory by exploiting the geometrical superfield formulation 
proposed by Bonora, {\it etal.} [14,15]. It should be noted that, in earlier attempts [7,13], the proper set of BRST 
and anti-BRST symmetries have {\it not} been obtained {\it together}. We have accomplished the goal
of obtaining the proper set of (anti-)BRST symmetry transformations by exploiting the superfield formalism.
One of the key consequences of this geometrical superfield approach (to BRST formalism) 
is a very natural derivation of the (anti-)BRST invariant 
CF-type restrictions that are found to be responsible for

(i) the absolute anticommutativity property of the off-shell 
(anti-)BRST symmetries in our present theory, and

(ii) the derivation of the coupled Lagrangian densities 
that respect the above off-shell nilpotent and absolutely anticommuting  symmetries {\it together}.

It is very interesting to point out that when we merge the above two {\it proper} 
(i.e. off-shell nilpotent and absolutely anticommuting) (anti-)BRST symmetry transformations, we
observe that this combination could be made off-shell nilpotent but the resulting ``merged'' (anti-)BRST
transformations turn out to be {\it not} absolutely anticommuting in nature. At present, we do not have
any clue as to how to obtain the absolute anticommutativity between the ``merged'' BRST and anti-BRST
symmetry transformations. 
We feel strongly that a single gauge-invariant restriction on the (super)fields
would lead to the derivation of the absolutely anticommuting ``merged'' (anti-)BRST symmetry
transformations within the framework of geometrical superfield formalism. This procedure would also
entail upon the theory a set of ``merged'' (anti-)BRST invariant CF-type restrictions that would
be responsible for the accomplishment of the  absolute anticommutativity property of the ``merged'' (anti-)BRST symmetry
transformations. 
We hope to address this problem in the future.

It would be a very nice venture to compute all the conserved symmetry generators for the various 
continuous symmetry 
transformations that are possibly present in the theory. In this connection, it is gratifying
to point out that, in our recent couple of papers [23,24], we have derived the conserved charges
corresponding to continuous symmetry transformations and shown the existence
of some cute novel features, hitherto, unseen in the application of BRST
approach to some of the (non-)Abelian $p$-form gauge theories. To provide the geometrical 
interpretation for the CF-type restrictions in the language of the geometrical object called gerbes 
(see, e.g. [19,20] for details) is yet another future direction for further investigation. 
It would be also very fruitful endeavor to tap the 
potential and power of other superfield approaches to BRST formalism [25-27] that might turn out to be useful
in the derivation of the {\it proper} merged (anti-)BRST symmetry transformations for the theory. In the light
of a recent work [28] on the phenomenological implication of the topologically massive gauge theory, it would
be interesting to study its relevance in various kinds of processes that are allowed by the standard
model of particle physics.
We are deeply involved with the above
cited issues and our results would be reported elsewhere [29].\\

\noindent
{\bf Acknowledgements:}
Financial support from DST, Government of India,  
under the SERC project sanction grant No: - SR/S2/HEP-23/2006, is gratefully acknowledged.

\begin{center}
{\section {Appendix}}
\end{center}

We capture here the no-go theorem [10], within the framework of the  superfield formalism, 
which reconfirms the impossibility  of straightforward generalization of  the 4D topological 
massive {\it Abelian } theory to its {\it  non-Abelian } counterpart. Towards this end in mind,
let us begin with the following horizontality condition [without taking into account the presence
of the auxiliary 1-form field $K_\mu \equiv (K_\mu \cdot T)$]:
\begin{eqnarray}
 &\tilde {d} \tilde {\cal B}^{(2)}+ i\;\bigl (\tilde {\cal A}^{(1)}_{(h)} \wedge \tilde {\cal B}^{(2)} 
- \tilde {\cal B}^{(2)} \wedge \tilde {\cal A}^{(1)}_{(h)} \bigr ) = dB^{(2)} + i\;
\bigl (A^{(1)} \wedge B^{(2)}
- B^{(2)} \wedge A^{(1)} \bigr ), &
\end{eqnarray}
where r.h.s. implies 
$H^{(3)}= dB^{(2)} + i\;(A^{(1)} \wedge B^{(2)} - B^{(2)} \wedge A^{(1)})
=\frac{1}{3!} (dx^{\mu} \wedge dx^{\nu} \wedge dx^{\eta}) H_{\mu\nu\eta}$ with
the totally antisymmetric curvature tensor $H_{\mu\nu\eta} = (\partial_\mu B_{\nu\eta} 
+ \partial_\nu B_{\eta\mu} + \partial_\eta B_{\mu\nu})- \bigl [A_\mu \times B_{\nu\eta} + A_\nu \times B_{\eta\mu} + A_\eta \times B_{\mu\nu} \bigr ]$ (without the presence of the auxliary $K_\mu$ field). In the l.h.s.,
the 1-form super connection with the subscript $(h)$, namely;
\begin{eqnarray}
\tilde {\cal A}^{(1)}_{(h)}(x, \theta,\bar\theta ) = dx^\mu \;
{\tilde {\cal B}}_\mu^{(h)}(x, \theta,\bar\theta ) + d \theta \;{\tilde {\bar {\cal F}}}^{(h)}(x, \theta,\bar\theta ) + d \bar \theta\;
{\tilde {\cal F}}^{(h)} (x, \theta,\bar\theta ), 
\end{eqnarray}
is defined in terms of  the superfields obtained after the application of horizontalilty condition in equation (8).
One of the kinetic terms of the theory (i.e. $\frac{1}{12} H^{\mu\nu\eta} \cdot H_{\mu\nu\eta}$) would be defined, 
now, in terms of the above totally antisymmetric curvature tensor $H_{\mu\nu\eta}$.

We shall check whether the above kinetic term remains invariant under the (anti-)BRST
symmetry transformations that ensue from (76). Further, we shall test the sanctity of 
the ensuing (anti-)BRST symmetry transformations by the requirements of off-shell nilpotency and 
absolute anticommutativity which are connected with the basic tenets of BRST formalism.

Let us focus on the l.h.s. of (76). This contains wedge products of the following differentials which include
the Grassmannian as well as spacetime coordinates, namely;
\begin{eqnarray}
&&(dx^{\mu} \wedge dx^{\nu} \wedge dx^{\eta}),(dx^{\mu} \wedge dx^{\nu} \wedge d\theta ), 
(dx^{\mu} \wedge dx^{\nu} \wedge d \bar\theta ),
(dx^{\mu} \wedge d\theta \wedge d\theta ), (dx^{\mu} \wedge d\theta \wedge d\bar \theta ),\nonumber\\ 
&&(dx^{\mu} \wedge d\bar\theta \wedge d\bar \theta ), (d\theta \wedge d\theta \wedge d \theta ), 
(d\theta \wedge d\theta \wedge d\bar\theta ), (d\theta \wedge d\bar\theta \wedge d \bar\theta ),
(d\bar\theta \wedge d\bar\theta \wedge d \bar\theta ).
\end{eqnarray}
It can be noted, from the HC (76), that all the coefficients of the differentials  with Grassmannian 
variables have to be set equal to zero. There are {\it nine} such wedge
products as is evident from (78). The setting of these coefficients equal to zero leads to
\begin{eqnarray}
&& {\partial}_{\theta} {\tilde {\cal B}_{\mu\nu}} + {\partial}_{\mu} {\tilde  {\bar {\cal F_{\nu}}}}
-{\partial}_{\nu} {\tilde  {\bar {\cal F_{\mu}}}} -i [ {\tilde {\cal B}_{\mu\nu}} , {\tilde {\bar {\cal F}}}^{(h)} ] 
+ i [ {\tilde {\cal B}}^{(h)}_\mu  ,{\tilde   {\bar {\cal F_{\nu}}}}]
- i [ {\tilde {\cal B}}^{(h)}_\nu ,{\tilde   {\bar {\cal F_{\mu}}}}]=0, \nonumber\\
&& {\partial}_{\bar \theta} {\tilde {\cal B}_{\mu\nu}} + {\partial}_{\mu} {\tilde   {\cal F_{\nu}}}
-{\partial}_{\nu} {\tilde   {\cal F_{\mu}}} -i [ {\tilde {\cal B}_{\mu\nu}} , {\tilde  {\cal F}}^{(h)} ] 
+ i [ {\tilde {\cal B}}^{(h)}_\mu ,{\tilde    {\cal F_{\nu}}}] 
- i [ {\tilde {\cal B}}^{(h)}_\nu ,{\tilde    {\cal F_{\mu}}}]=0, \nonumber\\
&& {\partial}_{\mu} {\tilde {\cal \phi}} + {\partial}_\theta {\tilde  {\cal F_{\mu}}} 
+ {\partial}_{\bar \theta} {\tilde {\bar{\cal F_{\mu}}}}
+ i [ {\tilde {\cal B}}^{(h)}_\mu , {\tilde {\cal \phi}} ] 
+ i \;\{{\tilde { \bar {\cal F}}}^{(h)}, {\tilde   {\cal F_{\mu}}}\} + i \;\{{\tilde  {\cal F}}^{(h)}, {\tilde   {\bar {\cal F_{\mu}}}}\}=0, \nonumber\\
&& {\partial_{\bar\theta }{\tilde  {\cal \beta}}} + i [{\tilde  {\cal F}}^{(h)}, {\tilde  {\cal \beta}}]= 0, \quad
{\partial_\theta {\tilde  {\cal \phi}}} + {\partial_{\bar\theta} {\tilde {\bar {\cal \beta}}}}
+ i \; [{\tilde {\bar {\cal F}}}^{(h)}, {\tilde {\cal \phi}}] + i \;[{\tilde {\cal F}}^{(h)}, {\tilde {\bar {\cal \beta}}}]= 0, 
\nonumber\\  
&& {\partial_\theta {\tilde {\bar {\cal \beta}}}} + i [{\tilde {\bar {\cal F}}}^{(h)}, {\tilde {\bar {\cal \beta}}}]= 0,
\quad {\partial}_{\bar \theta} {\tilde   {\cal F_{\mu}}} + {\partial}_{\mu} {\tilde  {\cal \beta}}
+ i \; [{\tilde  {\cal B}}^{(h)}_\mu, {\tilde  {\cal \beta}}] 
+ i \;\{{\tilde {\cal F}}^{(h)}, {\tilde  {\cal F_{\mu}}}\}= 0, \nonumber\\ 
&& {\partial}_\theta {\tilde  {\bar {\cal F_{\mu}}}} + {\partial_{\mu} {\tilde {\bar {\cal \beta}}}}
+ i \; [{\tilde  {\cal B}}^{(h)}_\mu, {\tilde {\bar {\cal \beta}}}] 
+ i \;\{{\tilde {\bar {\cal F}}}^{(h)}, {\tilde {\bar {\cal F_{\mu}}}}\}= 0, \nonumber\\
&& {\partial_{\bar\theta} {\tilde  {\cal \phi}}} + {\partial_{\theta} {\tilde  {\cal \beta}}}
+ i \; [{\tilde  {\cal F}}^{(h)}, {\tilde {\cal \phi}}] + i \;[{\tilde{\bar {\cal F}}}^{(h)}, {\tilde {\cal \beta}}]= 0.
\end{eqnarray} 
In the above computation, we have taken into account the full-fledged expansions of (15)
and (23). It is worthwhile to point out that all the superfields (in the above) with superscript $(h)$ are the
superfields obtained after the application of HC (cf. equation (8) in Sec. III). 
Furthermore, we would like to emphasize that the HC in (76) entails upon the fact that, ultimately, the
kinetic term ($\frac{1}{12} H^{\mu\nu\eta} \cdot H_{\mu\nu\eta} $) {\it should}
remain independent of the Grassmannian  variables $\theta$ and $\bar\theta$. In other words, we demand
that the kinetic term should remain gauge (as well as (anti-)BRST) invariant quantity
in the theory. This requirement, as is well-known, is the universal feature of any arbitrary $p$-form gauge theory
under gauge transformations.

The above restrictions lead to the derivation of the secondary fields of the expansions (in (15) and (23))
in terms of the basic fields and auxiliary fields of the theory as:
\begin{eqnarray}
&& R_{\mu\nu} = - (D_\mu C_\nu - D_\nu C_\mu ) + (C \times
 B_{\mu\nu}),\quad
\bar R_{\mu\nu} = - (D_\mu \bar C_\nu - D_\nu \bar C_\mu ) + 
(\bar C \times B_{\mu\nu}), \nonumber\\
&& S_{\mu\nu} = - i\;(D_\mu \bar B_\nu ^{(1)} - D_\nu \bar B_\mu^{(1)} ) + i\;(D_\mu \bar C \times C_\nu - D_\nu \bar C \times 
C_\mu ) - i\; (C \times \bar R_{\mu\nu})
- (\bar B \times B_{\mu\nu}) \nonumber\\  && \equiv  i\;(D_\mu B_\nu ^{(2)} - D_\nu B_\mu^{(2)} )
- i\;(D_\mu C \times \bar C_\nu - D_\nu C \times \bar C_\mu )
+ i\; (\bar C \times R_{\mu\nu}) +  (B \times B_{\mu\nu})\nonumber\\
&& B_\mu^{(1)} = - D_\mu \beta + (C \times C_\mu), \quad 
S_\mu = -i (D_\mu  \beta \times \bar C ) - i (C \times \bar B_\mu^{(1)})- (\bar B_1 \times C_\mu), \nonumber\\ && {\bar B_\mu}^{(2)} = - D_\mu \bar \beta + (\bar C \times \bar C_\mu), \quad \bar S_\mu = -i (D_\mu  \bar \beta \times \ C ) + i (\bar C \times B_\mu^{(2)})
+ (B_2 \times \bar C_\mu) \nonumber\\
&& \bar f_1 = -(\phi \times \bar C),\quad  f_1 = -(\phi \times C), \quad \bar f_2 = -(\beta \times \bar C),\quad f_2 = -(\beta \times C),\nonumber\\ &&\bar f_3 = -(\bar \beta 
 \times \bar C),\quad  f_3 = -(\bar \beta \times C),\quad
 b_1= i [(\phi \times \bar C) \times C] - (\bar B_1 \times \phi), 
 \nonumber\\ 
&& b_3= i [(\bar \beta \times \bar C) \times C] -(\bar B_1 \times \bar \beta ), 
\quad B_\mu^{(2)} + {\bar B}_\mu^{(1)} = -D_\mu \phi + ({\bar C} \times {C_\mu} ) + ( C \times {\bar C}_\mu), \nonumber\\
&&( F_{\mu\nu} \times \beta ) = 0, \; ( F_{\mu\nu} \times {\bar \beta }) = 0, \;( F_{\mu\nu} \times \phi )= 0,
\; b_2= i [(\beta \times \bar C) \times C] - (\bar B_1 \times \beta). 
\end{eqnarray}
There are other relationships as well. However, they are found to be consistent with the ones
that are listed above.
If we identify ${\bar B}_1 = \bar B, \;B_2 = B,\; B_\mu^{(2)}= B_\mu$,\;
${\bar B}_\mu^{(1)} = \bar B_\mu$ and substitute the above values explicitly in the expansions (15) and (23),
we obtain the following BRST and anti-BRST transformations (by taking the analogy with our exercise
in Sec. III and IV)
\begin{eqnarray}
&& s_{b}C = \frac {1}{2}(C \times C), \quad s_b {\bar C} =  i B, \quad s_b {\beta} = -(\beta \times C),
\quad s_b {\bar \beta} = -(\bar \beta \times C), \nonumber\\
&& s_b \phi = - (\phi \times C),\quad s_b C_\mu = - D_\mu \beta + (C \times C_\mu ), \quad
s_b {\bar C}_\mu = B_\mu, \quad s_b  [B_\mu, B] = 0, \nonumber\\
&& s_b{\bar B}_\mu = - (D_\mu {\beta} \times {\bar C } ) - i ({\bar B} \times {C_\mu} ) 
+ (C \times {\bar B}_\mu ), \quad s_b F_{\mu\nu} = - (F_{\mu\nu}\times C), \nonumber\\
&& s_b \bar B = - (\bar B \times C) \quad s_b B_{\mu\nu} = - (D_\mu C_\nu - D_\nu C_\mu) +  (C \times B_{\mu\nu}), 
\quad s_b A_\mu = D_\mu C, \nonumber\\
&& s_b H_{\mu\nu\eta} = -(H_{\mu\nu\eta} \times C) + 
(F_{\mu\nu} \times {C_\eta})
+ (F_{\nu\eta} \times {C_\mu}) + (F_{\eta\mu} \times {C_\nu}),
\end{eqnarray}
\begin{eqnarray}
&& s_{ab}{\bar C} = \frac {1}{2}({\bar C} \times {\bar C}), \quad s_{ab} C =  i {\bar B}, 
\quad s_{ab} {\beta} = -(\beta \times {\bar C}),
\quad s_{ab} {\bar \beta} = -(\bar \beta \times \bar C), \nonumber\\
&& s_{ab} \phi = - (\phi \times \bar C),\quad s_{ab} {\bar C}_\mu = - D_\mu {\bar \beta} 
+ ({\bar C} \times {\bar C}_\mu ), \quad
s_{ab}  C_\mu = {\bar B}_\mu, \quad s_{ab} [{\bar B}_\mu, \bar B] = 0, \nonumber\\
&& s_{ab} B_\mu = - (D_\mu {\bar \beta} \times  C ) 
- i ( B \times {\bar C}_\mu) 
+ ({\bar C} \times  B_\mu), \quad s_{ab}F_{\mu\nu} 
= - (F_{\mu\nu}\times {\bar C}), \nonumber\\
&& s_{ab} B = - (B \times \bar C), \quad s_{ab}B_{\mu\nu} = - (D_\mu {\bar C}_\nu - D_\nu {\bar C}_\mu)+ (\bar C \times B_{\mu\nu}),
\quad s_{ab} A_\mu = D_{\mu} {\bar C}, \nonumber\\
&& s_{ab} H_{\mu\nu\eta} = -(H_{\mu\nu\eta} \times {\bar C}) + ( F_{\mu\nu} \times {{\bar C}_{\eta}} )
+ ( F_{\nu\eta} \times {{\bar C}_{\mu }}) + ( F_{\eta\mu} \times {{\bar C}_{\nu}} ).
\end{eqnarray}
Furthermore, our geometrical superfield formulation leads to the derivation of 
specific relationships amongst some of the secondary fields of the super expansions in equations (15) and (23). These
are nothing but the following (anti-)BRST invariant Curci-Ferrari
(CF) type restrictions (plus other (anti-)BRST invariant conditions) on our theory:
\begin{eqnarray}
&& B + {\bar B} = -i (C \times {\bar C}), \quad ( F_{\mu\nu} \times \beta ) = 0, \quad ( F_{\mu\nu} \times {\bar \beta }) = 0,  \nonumber\\
&& B_\mu + {\bar B}_\mu = -D_\mu \phi
+ ({\bar C} \times {C_\mu} ) + ( C \times {\bar C}_\mu ),
\quad ( F_{\mu\nu} \times \phi )= 0. 
\end{eqnarray}
It is important to point out that these CF-type restrictions play very crucial roles
as will become clear in our further discussions. As a side remark,
we would like to state that the transformations $s_b H_{\mu\nu\eta}$ and $s_{ab} H_{\mu\nu\eta}$
in (81) and (82) have been derived from (76) by equating the coefficients of $(dx^\mu \wedge dx^\nu \wedge dx^\eta$)
from the l.h.s and r.h.s.

At this stage, a few comments are in order. First, the CF condition $ B + \bar B = - i (C \times \bar C)$
has been derived from setting the coefficient of differential ($dx^\mu \wedge d\theta \wedge d\bar\theta$)
equal to zero (cf. Sec. III) in the context of superfield approach to ``scalar'' gauge symmetry. Second, the 
CF-type constraint
$(F_{\mu\nu} \times \bar\beta) = 0$ is obtained when we set equal to zero the coefficient of differential
($dx^\mu \wedge dx^\nu \wedge d\theta$). Similarly, the restriction $(F_{\mu\nu} \times \beta) = 0$ arises
from the setting the coefficient of ($dx^\mu \wedge dx^\nu \wedge d \bar\theta$)
equal to zero. Third, the CF-type restriction
$B_\mu + {\bar B}_\mu = -D_\mu \phi + ({\bar C} \times {C_\mu} ) + ( C \times {\bar C}_\mu)$ emerges from setting
the coefficient of differential $(dx^\mu \wedge d\theta \wedge d\bar \theta$)
equal to zero in the context of superfield
approach to ``vector'' gauge symmetry. Finally, the CF-type constraint $(F_{\mu\nu} \times \phi) = 0$ is found
when we equate the two expressions for $S_{\mu\nu}$ (as illustrated in equation (80)) and perform
some algebraic simplifications.

Normally, our geometrical superfield approach leads to the derivation of the off-shell nilpotent (anti-)BRST symmetry
transformations (as is evident from Sec. III and Sec. IV). However, the (anti-)BRST symmetry transformations
emerging from the HC (76) have some special features. For instance, it can be checked that $s_b^2 B_{\mu\nu} 
= 0$ only when we exploit the CF-type restriction $(F_{\mu\nu} \times \beta) = 0$. Similarly, we notice
that $s_{ab}^2\; B_{\mu\nu} = 0$ only when we tap the power of the CF-type restriction 
$(F_{\mu\nu} \times \bar\beta) = 0$. Added to the above observations, it is interesting to point out that
$s_b^2 H_{\mu\nu\eta} = 0$ when we exploit the relations $(F_{\mu\nu} \times \beta) = 0$ and the
Bianchi identity (i.e. $D_\mu F_{\nu\eta} + D_\nu F_{\eta\mu} + D_\eta F_{\mu\nu} = 0$). Similar is the
situation with $s_{ab}^2 H_{\mu\nu\eta} = 0$ because we have to use $(F_{\mu\nu} \times \bar\beta) = 0$ and
$D_\mu F_{\nu\eta} + D_\nu F_{\eta\mu} + D_\eta F_{\mu\nu} = 0$. It is worthwhile to state,
in passing, that the off-shell
nilpotency of the BRST and anti-BRST symmetry transformations is valid 
very naturally for all the rest of the fields of the theory.

Now a few comments follow on the absolute anticommutativity property of the nilpotent (anti-)BRST symmetry transformations
that have emerged from the HC (76). First, as is well known, the CF condition $B + \bar B + i (C \times \bar C) = 0$
is used for the absolute anticommutativity property $\{ s_b, s_{ab} \} A_\mu = 0$. Second, it can be checked
that the above CF condition is used in the proof of $\{s_b, s_{ab} \} \beta = 0, \{s_b, s_{ab} \} \bar\beta = 0$.
Third, it is interesting to point out that $\{ s_b, s_{ab} \} B_{\mu\nu} = 0$ only when the above CF condition
as well as the CF-type restrictions
$B_\mu +  \bar B_\mu + D_\mu \phi - (C \times \bar C_\mu )- (\bar C \times C_\mu) = 0$ and
$(F_{\mu\nu} \times \phi) = 0$ are exploited together. Finally, it can be explicitly checked that
$\{ s_b, s_{ab} \} \;H_{\mu\nu\eta} = 0$ only when we exploit the CF-type restrictions
$B + {\bar B} = -i (C \times {\bar C}), B_\mu + {\bar B}_\mu = -D_\mu \phi
+ ({\bar C} \times {C_\mu} ) + ( C \times {\bar C}_\mu), (F_{\mu\nu} \times \phi) = 0$ together
with the validity of the Bianchi identity $D_\mu F_{\nu\eta} + D_\nu F_{\eta\mu} + D_\eta F_{\mu\nu} = 0$.

Finally, as pointed out very clearly earlier,  the real acid-test of the 
sanctity of the HC in (76) is the invariance of the kinetic term ($\frac{1}{12} H^{\mu\nu\eta} \cdot H_{\mu\nu\eta}$)
under the nilpotent (anti-)BRST symmetry transformations.
Thus, we have to clearly check the (anti-)BRST invariance of the kinetic
term (i.e. $\frac{1}{12} H^{\mu\nu\eta} \cdot H_{\mu\nu\eta}$) of the Lagrangian density of 
our present theory.
In this respect, it can be seen, from the continuous
transformations in (81) and (82), that the kinetic term
does {\it not} remain invariant under the (anti-)BRST transformations. As a consequence, the 
no-go theorems, proposed in [10-12], are {\it correct} because the straightforward generalization of the
topologically massive {\it Abelian} theory to its {\it non-Abelian} counterpart (under the purview of the HC in (76))
lands us in difficulties. This is due to the fact that (i) the (anti-)BRST transformations are highly constrained
(cf. (83)), and (ii) the kinetic term does {\it not} remain invariant under the (anti-)BRST transformations
(82) and (81).

\end{document}